\documentclass[aps,prx,twocolumn,superscriptaddress,nofootinbib]{revtex4-2}

\usepackage{amssymb}
\usepackage{amsmath}
\usepackage{bbold}
\newtheorem{theorem}{Theorem}
\usepackage{braket}
\usepackage{xcolor}
\usepackage{soul}
\usepackage{graphicx}
\usepackage{textcomp}
\graphicspath{{../figura1/}{../figura2/}}
\newcommand*\diff{\mathop{}\!\mathrm{d}}

\definecolor{hotpink}{rgb}{1.0, 0.41, 0.71}

\DeclareMathOperator{\Tr}{Tr}
\DeclareMathOperator{\He}{He}
\DeclareMathOperator{\id}{id}
\DeclareMathOperator{\expect}{\mathbb{E}}
\DeclareMathOperator{\diag}{diag}

\usepackage[breaklinks]{hyperref}
\hypersetup{
    colorlinks,
    linkcolor={red!80!black},
    citecolor={green!70!black},
    urlcolor={blue!80!black}
}

\usepackage{xcite}
\usepackage{xr}

\makeatletter
\newcommand*{\addFileDependency}[1]{
\typeout{(#1)}
\IfFileExists{#1}{}{\typeout{No file #1.}}
}
\makeatother

\begin{document}


\title{A statistical mechanics framework for Bayesian deep neural networks beyond the infinite-width limit}

\author{R. Pacelli}
\affiliation{Dipartimento di Scienza Applicata e Tecnologia, Politecnico di Torino, 10129 Torino, Italy}
\affiliation{Artificial Intelligence Lab, Bocconi University, 20136 Milano, Italy}

\author{S. Ariosto}
\affiliation{Dipartimento di Scienza e Alta Tecnologia and Center for Nonlinear and Complex Systems,
Università degli Studi dell'Insubria, Via Valleggio 11, 22100 Como, Italy}
\affiliation{I.N.F.N. Sezione di Milano, Via Celoria 16, 20133 Milano, Italy}

\author{M. Pastore}
\affiliation{Universit\'{e} Paris-Saclay, CNRS, LPTMS, 91405 Orsay, France}
\affiliation{Laboratoire de physique de l'\'{E}cole normale sup\'{e}rieure, CNRS, PSL University,
Sorbonne University, Universit\'{e} Paris-Cit\'{e}, 24 rue Lhomond, 75005 Paris, France}

\author{F. Ginelli}
\affiliation{Dipartimento di Scienza e Alta Tecnologia and Center for Nonlinear and Complex Systems,
Università degli Studi dell'Insubria, Via Valleggio 11, 22100 Como, Italy}
\affiliation{I.N.F.N. Sezione di Milano, Via Celoria 16, 20133 Milano, Italy}

\author{M. Gherardi}
\affiliation{Università degli Studi di Milano, Via Celoria 16, 20133 Milano, Italy}
\affiliation{I.N.F.N. Sezione di Milano, Via Celoria 16, 20133 Milano, Italy}

\author{P. Rotondo}
\affiliation{Dipartimento di Scienze Matematiche, Fisiche e Informatiche,
Università degli Studi di Parma, Parco Area delle Scienze, 7/A 43124 Parma, Italy}
\affiliation{I.N.F.N. Sezione di Milano, Via Celoria 16, 20133 Milano, Italy}

\begin{abstract}
Despite the practical success of deep neural networks, a comprehensive theoretical framework that can predict practically relevant scores, such as the test accuracy, from knowledge of the training data is currently lacking. Huge simplifications arise in the infinite-width limit, where the number of units $N_\ell$ in each hidden layer ($\ell=1,\dots, L$, being $L$ the depth of the network) far exceeds the number $P$ of training examples. This idealisation, however, blatantly departs from the reality of deep learning practice. Here, we use the toolset of statistical mechanics to overcome these limitations and derive an approximate partition function for fully-connected deep neural architectures, which encodes information about the trained models. The computation holds in the ``thermodynamic limit'' where both $N_\ell$ and $P$ are large and their ratio $\alpha_\ell = P/N_\ell$ is finite. This advance allows us to obtain
(i) a closed formula for the generalisation error associated to a regression task in a one-hidden layer network with finite $\alpha_1$;
(ii) an approximate expression of the partition function for deep architectures (via an ``effective action'' that depends on a finite number of ``order parameters'');
(iii) a link between deep neural networks in the proportional asymptotic limit and Student's $t$ processes.
\end{abstract}

\maketitle

\section{Introduction}

The rise of deep learning, driven by advances in computing technology and foreshadowed by decades of research, has outpaced our ability to develop a solid theoretical foundation
\cite{GoodfellowBook,EngelVanDenBroeck}.
Filling the gaps in our understanding of deep learning on a fundamental level
is a long-time collective effort involving several communities.
Statistical physics achieved far-reaching results in this regard,
and remains a wellspring of fresh perspectives and breakthroughs
\cite{
seroussi2023natcomm,
Wakhloo:2023,
SompolinskyLinear,
BaldassiLauditi:2022,
canatar2021,
Saad:PRL:2020,
Goldt:PRX:2020,
doi:10.1146/annurev-conmatphys-031119-050745,
Saad:PRL:2018}.
One notable recent advance was obtained by considering the infinite-width limit, where the number of training data $P$ is fixed and the size of the hidden layers is taken to infinity. The observation that such deep models are equivalent to Gaussian processes (GPs)~\cite{Neal,NIPS1996_ae5e3ce4,g.2018gaussian, LeeGaussian, garriga-alonso2018deep, novak2019bayesian, JacotNTK, ChizatLazy, lee2019wide} established a connection between deep learning and kernel methods  \cite{cortes1995support}, and provided a statistical physics description of this regime~\cite{pmlr-v119-bordelon20a, canatar2021, PhysRevLett.82.2975}.

However, there is agreement that a more complete theory should address deep learning beyond the infinite-width limit~\cite{Seleznova2022, Vyas2022,antognini2019finite,yaida2020nonGauss,hanin2023random,zavatone2021prior}: in fact, realistic neural networks operate in a qualitatively different regime, where the number of training examples exceeds 
the width of the largest layer. 
Modeling the finite-width regime in 
the thermodynamic limit, where the number of degrees of freedom diverges and the tools of statistical mechanics are most effective,
amounts to taking the asymptotic limit where both the size of the training set $P$ and the number of units in each hidden layer $N_\ell$ are taken to infinity with their ratios fixed, as we consider in the present work:  
\begin{equation}
    P, N_\ell \to \infty, \quad \alpha_\ell = \frac{P}{N_\ell} \,\, \text{finite} \qquad \forall \ell = 1, \ldots, L
    \label{eq:finite_scaling}
\end{equation}
with $L$ being the (finite) depth of the network (the scaling of $P$ with the input size $N_0$, which deserves special care, is discussed in section \ref{sec:hypothesis} of the Methods). This choice guarantees that such networks work in the overparametrised regime. 

Another fruitful line of research, in the direction of overcoming the limitations of the infinite-width limit,
sacrifices the non-linear nature of the network by considering a deep \emph{linear} input-output mapping: even if the resulting architecture lacks the expressive power~\cite{bengio2011expressive, bartlett2019nearly, PhysRevResearch.2.023169, PhysRevLett.125.120601, PhysRevE.102.032119,GherardiEntropy,Pastore_2021,Aguirre2022} of the same model with non-linearities, the multi-layer structure maintains the learning problem non-convex, while amenable to analytical investigation~\cite{Saxelinear, doi:10.1073/pnas.1820226116}. Very recently, 
Li and Sompolinsky~\cite{SompolinskyLinear} proposed a method to analytically evaluate properties of deep finite-width linear networks 
(e.g., their generalisation error) 
trained on a generic fixed training set. However, the more relevant case of generic non-linear DNNs 
remains an open problem, despite some recent notable attempts to address it~\cite{NEURIPS2021_cf9dc5e4, NEURIPS2021_b24d2101, seroussi2023natcomm, PhysRevE.105.064118, antognini2019finite, yaida2020nonGauss,hanin2023random, zavatone2021prior}. 

In statistical mechanics, the partition function is the central object encoding the properties of the system in the thermodynamic limit.
In this work we address the analytical computation of the partition function of a fully-connected, multi-layer, non-linear neural network,
as a function of the training set in the asymptotic limit defined in \eqref{eq:finite_scaling}.
Technically, the computation amounts to integrating out an extensive number of degrees of freedom (the weights of the network),
thus landing on an expression that involves only a finite number
(proportional to the depth $L$) of integrals,
to be evaluated by the saddle-point method.
In the one-hidden-layer (1HL) case, the only key approximation is justified by a generalised central limit theorem due to Bardet and Surgailis \cite{bardet2013} (which belongs to a class of results known as Breuer-Major (BM) theorems~\cite{NourdinQuantitative} from the seminal paper \cite{BM}).

In the general case of an architecture with $L$ hidden fully-connected layers, we show that the distribution of the pre-activations at each layer $\ell$ is a mixture of Gaussians that depends on $\ell$ parameters.
Notably, the back-propagating integration performed in \cite{SompolinskyLinear} is not a viable option as soon as non-linearities are added to the model. 
We introduce a forward-propagating method to carry out nested integrations starting from the input layer.
This result depends on an assumption that is similar, at least in spirit, to the Gaussian equivalence principle employed for random and generic feature models \cite{pmlr-v119-20a, loureiro2021learning, goldt2020gaussian, 10.2307/26542784, mei2019, 10.1214/20-AOS1990, Ariosto}. 

From these developments,
we are able to obtain quantitative predictions for the generalisation error of the network below the interpolation threshold. Moreover, our results have an intriguing interpretation from the point of view of stochastic processes: we show that in the case of finite $\alpha_\ell$ the GP arising in the infinite-width limit of Bayesian neural networks~\cite{LeeGaussian} should be generalised to a Student's $t$ stochastic process~\cite{Shah2014}.

As a first application
of the theory, we establish a simple criterion (equivalent to the one found in the linear case  \cite{SompolinskyLinear,zavatone2022contrasting} and in finite-$P$ perturbation theory  \cite{zavatone2021asymptotics,zavatone2022asymptoticsJStatMech}) to predict whether it is convenient, in terms of generalisation performance, to employ a finite-width deep neural network over its infinite-width version. 

\paragraph*{Problem setting -}
\label{sec:setting}
We consider a supervised learning problem with training set $\mathcal T_P = \{\mathbf x^\mu, y^\mu\}_{\mu=1}^P$, where each $\mathbf x^\mu \in \mathbb R^{N_0}$ and the corresponding labels $y^\mu \in \mathbb R$. The architecture is a deep neural network $f_{\textrm{DNN}} (\mathbf x)$ with $(L-1)$ fully-connected hidden (FC) layers and a final linear readout layer as defined in \eqref{eq:f_DNN}. We analyse regression problems with a quadratic loss function:
\begin{align}
\mathcal L &= \frac{1}{2} \sum_{\mu=1}^P \left[y^\mu - f_{\textrm{DNN}}(\mathbf x^\mu)\right]^2 + \mathcal L_{\textrm{reg}}\,,\\
\mathcal L_{\textrm{reg}} &= \frac{\lambda_L}{2\beta} \sum_{i_{L}=1}^{N_L} v_{i_L}^2 + \frac{1}{2\beta} \sum_{\ell=0}^{L-1} \lambda^{(\ell)} \lVert W^{(\ell)}\rVert^2\,
\label{loss},
\end{align}
where $L^2$ regularisations have been added for each layer to the loss function, $\lVert \cdot\rVert $ is the standard Frobenius norm defined for the weights matrices $W^{(\ell)}$, and $\beta =1/T$ is the inverse temperature parameter. 

As a standard practice in statistical mechanics of deep learning, we define the partition function of the problem as:
\begin{equation}
Z = \int \!\mathcal D \theta \,e^{-\beta \mathcal L (\theta)}\,.
\label{defpartition}
\end{equation}
where the symbol $\int \mathcal D \theta$ indicates the collective integration over the weights of the network, $\theta = \{ W^{(\ell)} , v  \}$. 
This choice enforces minimization of the training error for $\beta \to \infty$. We notice that scaling $\mathcal L_{\textrm{reg}}$ by $1/\beta$ has a natural Bayesian learning interpretation: the Gibbs probability $P_{\beta} (\theta)=Z^{-1} e^{-\beta \mathcal L (\theta)}$ associated with the partition function in equation \eqref{defpartition} is the posterior distribution of the weights after training, whereas the Gaussian regularization is a prior equivalent to assuming that weights at initialization have been drawn from a Gaussian distribution

In this framework, the average test error over a new (unseen) example $(\mathbf{x}^0, y^0)$ is given by:
\begin{equation}
\braket{\epsilon_{\textrm g} (\mathbf{x}^0, y^0)} = \int \!\mathcal D \theta \, [y^0- f_{\textrm{DNN}}(\mathbf{x}^0)]^2 \frac{e^{-\beta \mathcal L(\theta)}}{Z}\,.
\label{eq:err_g_def}
\end{equation}

\section{Results}

\begin{figure}[t!]
\includegraphics[width=0.45\textwidth]{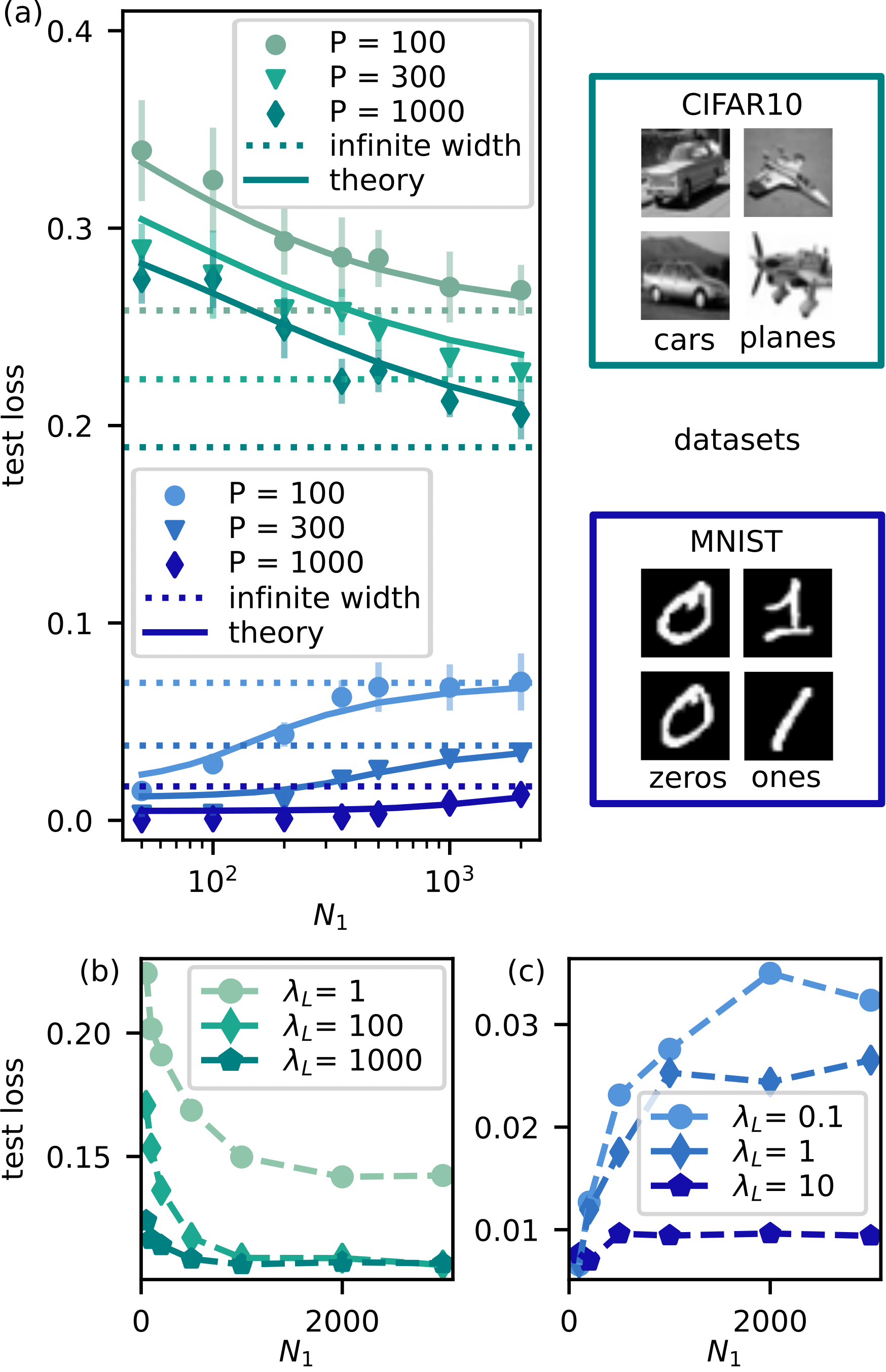}
\caption{(a) Learning curves of 1HL architectures with Erf activation (trained with a discretised Langevin dynamics, see also Methods) as a function of the hidden layer size $N_1$ for two regression tasks on the CIFAR10 (above) and MNIST (below) datasets. Zero/one labels have been chosen in both cases and the images of the CIFAR10 dataset have been gray-scaled and down-scaled to $N_0 = 28 \times 28$. The experimental test loss at different values of the trainset size $P$ (points with error bars indicating one standard deviation) are compared with the theory computed from equation \eqref{Eps_g} (solid lines). The bar centres are computed as the average over an ensemble of $S = 450$ equilibrium configurations. Samples are taken every $10^4$ Langevin steps (after thermalisation of the dynamics). The error bar represents one standard deviations from the average. (b,c) Experimental learning curves as a function of $N_1$ for increasing values of the Gaussian prior of the last layer $\lambda_1$. Error bars are within points, and dashed lines connecting the points are shown to guide the eye. The nets are trained on $P = 3000$ examples from the CIFAR10 dataset in (b) and $P = 500$ examples from MNIST in (c). Two qualitative predictions of the theory at zero temperature are checked: (i) the generalisation loss should decrease for any $N_1$ when $\lambda_1$ grows; (ii) the dependence of the learning curves on $N_1$ disappears in the large-$\lambda_1$ limit, since the bias is constant (see also main text).}
\label{fig:1}
\end{figure}

\subsection{Asymptotic effective action for one-hidden-layer neural networks in the Bayesian setting}
\label{sec:1HL}

In the case of 1HL architectures, we are able to reduce the partition function \eqref{defpartition} to the following two-variables integral in the thermodynamic limit  described in \eqref{eq:finite_scaling}:

\begin{equation}
Z = \int\diff Q \int \diff \bar{Q}\, \exp\left[{-\frac{N_1}{2} S(Q, \bar Q)}\right]
\label{FinalZ}
\end{equation}
where we have defined an effective action $S$ given by:
\begin{equation}
\begin{aligned}
S= &-Q\bar{Q}+\log(1+Q)+\frac{\alpha_1}{P}\Tr\log \beta\left[ \frac{ \mathbb{1}}{\beta} +\frac{ \bar{Q}K}{\lambda_1}\right]\\
&+\frac{\alpha_1}{P} y^\top \left[ \frac{\mathbb{1}}{\beta} +\frac{\bar{Q}K}{\lambda_1}\right]^{-1} y
\end{aligned}
\label{effS}
\end{equation}
and we have introduced a vectorial notation for the output $y^\top = (y^1,y^2,\dots, y^P)$. The $P \times P$, input-dependent kernel $K/\lambda_1$ is the neural network Gaussian process (NNGP) kernel \cite{LeeGaussian} arising in the infinite-width limit and its precise definition in terms of the input covariance matrix (rescaled by the Gaussian prior of the first layer $\lambda_0$) $C_{\mu\nu} = \mathbf{x}^\mu \cdot \mathbf{x}^\nu/(\lambda_0 N_0)$ is given in the Methods, equation~\eqref{methods:eq:kernel}. Note also that equation~\eqref{effS} holds for zero-mean activation functions, that is functions whose average over a centered Gaussian is zero (see equation~\eqref{methods:eq:zeromean}; an effective action for the generic finite-mean case is reported in the supplemental material \cite{supplemental}, Sec.~\ref{supp:sec:finitemeanactivation}) and that for many reasonable non-linearities and input data distributions the derivation goes through at least in the regime $P = O(N_0)$ (we discuss this key technical point in the Methods). This is the first main result of our work and we conjecture it is exact since the only key Gaussian approximation that we perform is justified by the extension of the Breuer-Major theorem~\cite{bardet2013}, as argued in the Methods. 

In the supplemental material \cite{supplemental}, we obtain a number of additional results that did not enter here for space limitations: (i) a re-derivation of the effective action in the case of linear activation function, valid at fixed $P,N_1,N_0$, together with a comparison with the results given in \cite{SompolinskyLinear, hanin2023bayesian}; (ii) a specific derivation of the effective action for quadratic activation function, which makes no use of the Breuer-Major theorem; (iii) the generalisation of the effective action in equation~\eqref{effS} to the case of multiple (but finite) outputs.

We can now solve equation~\eqref{effS} using the saddle-point method, since $N_1 \to \infty$, which amounts to finding the solutions $Q^*$, $\bar Q^*$ of the system of equations $\partial_Q S = 0$, $\partial_{\bar Q} S = 0$ (the infinite-width limit is re-obtained for $\alpha_1\to0$ and corresponds to the particular solution $Q^* = 0$, $\bar Q^* = 1$). In the zero-temperature limit, we can find the analytical solution of the saddle-point equations (see Methods). A straightforward computation shows that the generalisation error is given in terms of the usual bias-variance decomposition:
\begin{equation}
\begin{aligned}
&\braket{\epsilon_\text{g}(\mathbf{x}^0,y^0)} = \left(y^0 - \Gamma_1\right)^2 + \sigma_1^2\,,  \label{Eps_g} \\
&\quad\Gamma_1 = \sum_{\mu,\nu} \kappa_\mu(\mathbf x^0) K^{-1}_{\mu\nu}\; y_\nu\,,  \\
&\quad\sigma_1^2 =\frac{\bar Q^*}{\lambda_1} \left[\kappa_0(\mathbf x^0) - \sum_{\mu,\nu} \kappa_\mu(\mathbf x^0) K^{-1}_{\mu\nu}\; \kappa_\nu (\mathbf x^0)\right]\,,
\end{aligned}
\end{equation}
where $\kappa_{\mu} (\mathbf x^0)$, $\kappa_0 (\mathbf x^0)$ can be computed from the functional definition of the NNGP kernel using the new unseen input $\mathbf x^0$, as shown in the Methods.

We can directly employ equation~\eqref{Eps_g} to obtain testable predictions for the generalisation error of finite-width 1HL architectures trained in the Bayesian learning setting, as we do in panel (a) of Fig.~\ref{fig:1} for two specific regression tasks defined on the CIFAR10 and MNIST datasets (details on the numerical experiments are provided in the Methods section~\ref{sec:experiments} and in Sec. V of the supplemental material \cite{supplemental}). It turns out that the generalisation curves for the two regression tasks are monotonically increasing (decreasing) as a function of $N_1$ depending on the fact that the observable $y^\top \left(K/\lambda_1\right)^{-1} y/P$ is smaller (larger) than one. The importance of this quantity in controlling the generalisation performance has been already noted in linear networks \cite{SompolinskyLinear, zavatone2022contrasting} as well as in direct perturbation theory at finite $P$ for non-linear networks \cite{zavatone2021asymptotics,zavatone2022asymptoticsJStatMech}.

We also point out two semi-quantitative predictions for the general behavior of the generalisation error, just by looking at the dependence of equation \eqref{Eps_g} on the size of the hidden layer $N_1$ and on the Gaussian prior of the last layer $\lambda_1$. At $T=0$, the bias is constant as a function of $N_1$ (as explicitly observed also in the linear case in Ref.~\cite{SompolinskyLinear}) and of $\lambda_1$. On the contrary, the variance depends on $N_1$ and decreases as $1/\sqrt{\lambda_1}$ in the large-$\lambda_1$ limit. These observations lead to the following two testable predictions: (i) increasing the magnitude of the Gaussian prior $\lambda_1$ should systematically improve the generalisation performance at any $N_1$; (ii) for large $\lambda_1$ the dependence on $N_1$ of the generalisation error should disappear (see also the numerical experiments performed in panel (b) in Fig.~\ref{fig:1}).\\

\begin{figure*}[t!]
\includegraphics[width=.95\textwidth]{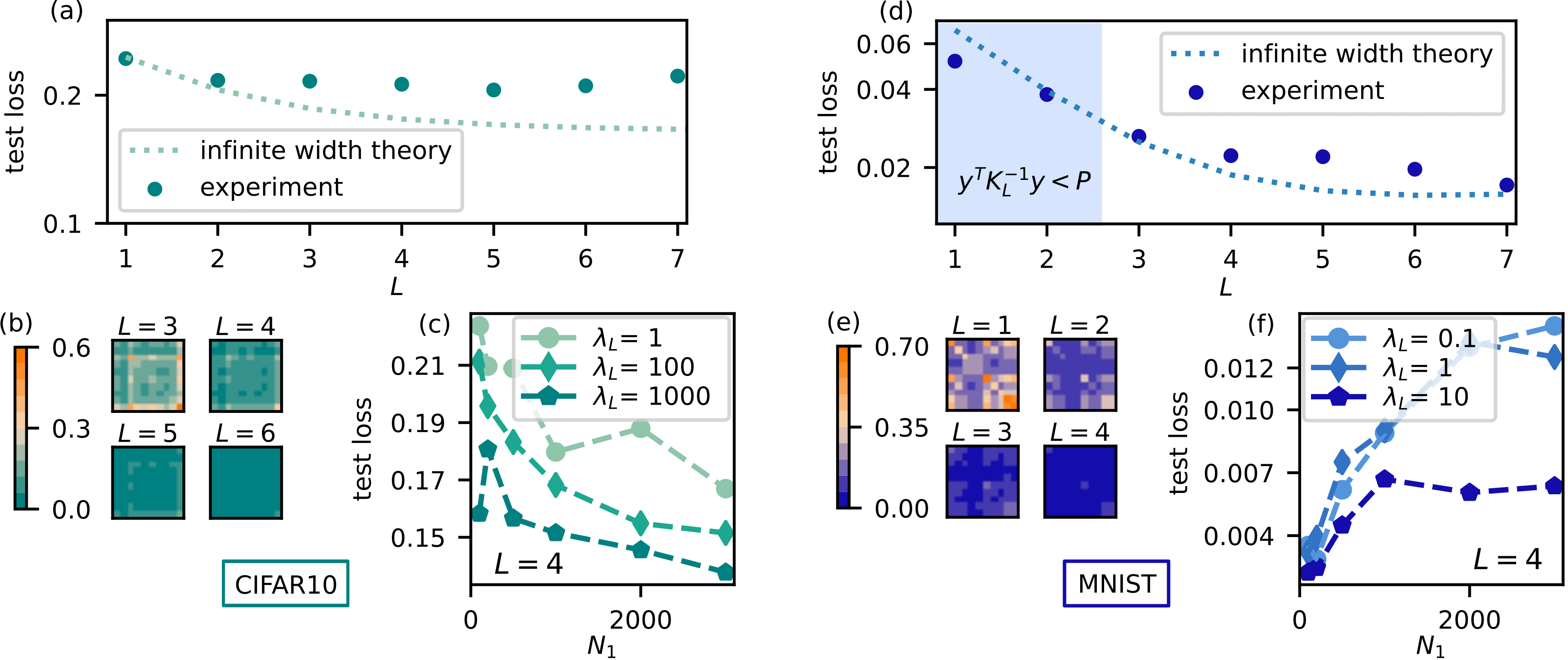}
\caption{ (a,d) Test loss of a $L$-HL neural network with ReLU activation, as a function of the depth $L$, for $P = 100$. The net is trained on a regression task in the small $\alpha$ regime ($\alpha = 0.1$), close to the infinite-width limit. The finite-width network can outperform the infinite-width prediction only when $s_L < 1$ (shaded area), i.e. only for the MNIST task and for depth $L<3$. (b,e) Visualisation of the entries of the infinite-width NNGP kernel at different layers of the network. The ReLU NNGP kernel converges to zero after repeated iterations. This generates almost vanishing eigenvalues that makes $s_L$ eventually always larger than one. (c,f) Test loss of a 4-HL network trained on $P= 1000$ examples with different regularisation strengths (with $N_\ell = N = 1000$). While increasing the magnitude of the Gaussian prior of the last layer still improves generalisation for all $N$, it is not clear anymore (as it was for 1HL networks) that the curve at large $\lambda_L$ is a constant as a function of $N$. The dashed line is shown to guide the eye. In all panels, error bars lie within points.}
\label{fig:2}
\end{figure*}

\subsection{Link between Student's \texorpdfstring{$t$}{t}-processes and shallow neural networks in the proportional limit}
\label{sec:student-t}

In obtaining the results reported in Sec.~\ref{sec:1HL}, our theory can be formulated as a statement on the probability distribution of the output variables
\begin{equation}
\begin{gathered}
    s^\mu \equiv \frac{1}{\sqrt{N_1}} \sum_{i_1=1}^{N_1} v_{i_1} \sigma(h_{i_1}^\mu)\,,\\
\end{gathered}
\label{eq:smu}
\end{equation}
where $h\sim\mathcal{N}(0,C\otimes \mathbb{1}_{N_1}) $, $v \sim \mathcal{N}(0,\lambda_1^{-1} \mathbb{1}_{N_1})$. Proceeding as in the derivation of the partition function presented in Methods, the p.d.f. of these variables can be written as a re-weighted Fourier transform,
\begin{equation}
    P(s|\mathcal{T}_P) = \frac{e^{-\frac{\beta}{2}\sum_{\mu} (y^\mu - s^\mu)^2}}{Z} \int \prod_\mu\frac{\diff \bar{s}^\mu}{2\pi}\,  e^{i \bar{s}^\top s} \Xi(\bar{ s}) \,,
    \label{eq:P(s|data)}
\end{equation}
of the function
\begin{equation}
	\Xi(\bar{ s}) = \left(1+ \frac{1}{\lambda_1 N_1}\sum_{\mu,\nu}^P \Bar{s}^\mu K_{\mu \nu}(C)\Bar{s}^\nu \right)^{-\frac{N_1}{2}}\,.
 \label{eq:Xi}
\end{equation}
It is straightforward to notice that as long as $N_1 \to \infty$ and $N_1 \gg P$, the dependence on $N_1$ disappears and we get:
\begin{equation}
\Xi(\bar{s}) \to e^{-\frac{1}{2\lambda_1} \sum_{\mu,\nu}^P \Bar{s}^\mu K_{\mu \nu}(C)\Bar{s}^\nu}\,.
	\label{eq:P(Q)largeN}
\end{equation}
This quantity has a very natural interpretation in view of the NNGP literature. Indeed, for $N_1$ large and $P$ finite, the variables~\eqref{eq:smu} are jointly multivariate Gaussian distributed according to the central limit theorem, as noted for example in~\cite{LeeGaussian}: this limit corresponds indeed to the RHS of our equation~\eqref{eq:P(Q)largeN} and is the cornerstone of the mapping of an infinite-width Bayesian neural network to a GP. This is however no more the case when $P$ is comparable to $N_1$: equation~\eqref{eq:Xi}, derived exploiting the Gaussian equivalence based on the BM theorem in the proportional asymptotic limit $P/N_1 \sim O(1)$, is suggesting that the variables $\bar{s}^\mu$ are distributed according to a multivariate Student's $t$-distribution~\cite{Shah2014,Coolen_2020,Coolen_pre,Uchiyama2021}.

The need of considering Student's $t$-processes as a generalisation of NNGPs has been noted already in the case of different priors on the distribution of the last layer's weights~\cite{Lee2022scale}. Non-Gaussianity of the posterior in a form similar to that of Eq. \eqref{eq:Xi} has appeared also in \cite{aitchison2020bigger,zavatone2021depth, yang2023theory}. The reason why this kind of process arises in the case we are considering here can be understood with an heuristic argument: when $N_1$ and $P$ are of the same order, we cannot take the limit $N_1\to \infty$ before $P\to \infty$, and so we need to use the empirical covariance of the output variables $s^\mu$ instead of their true one in estimating their probability distribution. A more precise characterization of these neural network Student's $t$-processes (NNTPs) and the regime where they arise represent interesting topics for future work.

\subsection{Asymptotic effective action for deep neural networks in the Bayesian setting}
\label{sec:LHL}

In the generic case of a deep fully-connected architecture with a finite number of layers $L$ and zero-mean activation function, we express the partition function in terms of a $2L$-dimensional integral (see Methods):
\begin{equation}
Z_{\textrm{DNN}} = \int \prod_{\ell =1}^{L} \diff Q_\ell \diff \bar Q_\ell e^{-\frac{N_L}{2} S_{\textrm{DNN}}(\{Q_\ell ,\bar Q_\ell\})}\,,
\end{equation}
where the effective action is given by:
\begin{align}
S_{\textrm{DNN}}&= \sum_{\ell=1}^{L} \frac{\alpha_L}{\alpha_\ell}\left[ -Q_\ell \bar Q_{\ell} + \log(1+Q_\ell)\right] \nonumber \\
&+\frac{\alpha_L}{P}\text{Tr}\log \beta \left(  \frac{\mathbb{1}}{\beta} +K^{(R)}_L( \{\bar Q_\ell\})\right) \notag\\
&+\frac{\alpha_L}{P} y^T \left( \frac{\mathbb{1}}{\beta} + K^{(R)}_L( \{\bar Q_\ell\})\right)^{-1} y\,
\end{align}
and we have introduced a renormalised kernel $K^{(R)}$ that generalises the the recurrence relation for the $L$-layer NNGP kernel as:
\begin{equation}
K^{(R)}_{\ell}(\{\bar Q_\ell\}) = \bar Q_{\ell}/\lambda_{\ell} K \circ\left[ K^{(R)}_{\ell-1}(\{\bar Q_\ell\}) \right]\,,\;\;\; K^{(R)}_0 = C\,,
\label{K_LQ}
\end{equation}
where $C$ is the covariance matrix of the inputs defined above and we stress that each $K^{(R)}_{\ell}$ depends on the variables $\bar Q_1,\dots,\bar Q_{\ell-1}$ only. For completeness, we notice that the recurrence relation for the infinite-width kernel $K_L$ is given by equation \eqref{K_LQ} with $\bar Q_\ell = 1$ $\forall \ell =1,\dots, L$.

This action shares the same structure as the one found in section \ref{sec:1HL} for the special case of 1HL, with the difference that for $L$ hidden layers, the recursive nature of the derivation introduces additional order parameters that are nested in the definition of the kernel $K_L$. Furthermore, since our derivation applies to layers of arbitrary size $N_\ell$, the action also depends on the aspect ratios $\alpha_\ell$. In the supplemental material \cite{supplemental}, we derive a series of additional results: (i) we generalise this effective action for finite-mean activation functions; (ii) we show how to recover the linear case in the isotropic limit $\alpha_\ell = \alpha $ $\forall \ell = 1,\dots, L $; (iii) using (i) we show how to correct the heuristic theory for ReLU activation presented in Ref. \cite{SompolinskyLinear}.

The computation of the generalisation error over a new example $(\mathbf x^0, y^0)$ gives:
\begin{equation}
\begin{split}
\braket{\epsilon_\text{g}(\mathbf{x}^0,y^0)} &= (y^0-\Gamma_L)^2+\sigma^2_L
\end{split}
\label{eq:gen_err0deep}
\end{equation}
where
\begin{align}
\Gamma_L &= \sum_{\mu\nu}\kappa^{(R)}_{L\mu} \left(\frac{\mathbb{1}}{\beta} + K^{(R)}_L( \{\bar Q_\ell\})\right)^{-1}_{\mu\nu} y_\nu, \label{GL}\\
\sigma^2_L &= \kappa^{(R)}_{L0} -\sum_{\mu\nu}\kappa^{(R)}_{L\mu} \left(\frac{\mathbb{1}}{\beta} + K^{(R)}_L( \{\bar Q_\ell\})\right)^{-1}_{\mu\nu}\kappa^{(R)}_{L\nu}\label{sigmaL}
\end{align}
and $\kappa^{(R)}_{L\mu}$, $\kappa^{(R)}_{L0}$ are recursive kernels computed from the recurrence given in equation \eqref{K_LQ} using the input $\mathbf x^0$ in the initial conditions. 

Note that also in this case we can perform the same scaling analysis of the dependence of the generalisation error on the Gaussian prior in the last layer $\lambda_L$ (in the zero temperature limit). It turns out that the bias does not depend on it, whereas the variance $\sigma_L^2$ approaches zero as $1/\sqrt \lambda_L$ as $\lambda_L$ is taken to infinity. This means that also in the case of finite depth $L > 1$, training at large values of the Gaussian prior of the last layer should improve generalisation at any aspect ratio of the network. We confirm this general observation with numerical experiments in panels (c) and (f) of Fig. \ref{fig:2}. However, differently from the 1HL case, we observe that the bias does depend on the aspect ratio even in the zero-temperature limit and we cannot expect anymore that the dependence on the aspect ratios of the networks $\alpha_\ell$ disappears in the $\lambda_L \to \infty$ limit.   

We can obtain another prediction of the theory at $L$ layers (that again confirms previous results on linear networks and perturbative calculations for non-linear networks \cite{SompolinskyLinear,zavatone2022contrasting, zavatone2021asymptotics,zavatone2022asymptoticsJStatMech}) by considering the effective action for ReLU activation. A straightforward Taylor expansion around the infinite-width limit $\alpha_\ell = \alpha = 0$ $\forall \ell=1,\dots,L$ shows that the first correction to the test loss $\Delta \epsilon_{\textrm g}$ is proportional to:
\begin{equation}
\Delta \epsilon_{\textrm g} \propto \alpha \left(\frac{1}{P}y^T K_L^{-1} y -1\right)\,.
\label{correction}
\end{equation}
where $K_L$ is the solution of recurrence in equation \eqref{K_LQ} for $\bar Q_\ell = 1$ $\forall \ell=1,\dots, L$ and ReLU activation. This means that  there exists a simple scalar observable that determines whether the finite-width deep neural network will outperform its infinite-width counterpart that generalises the one found at 1HL:
\begin{equation}
s_L = \frac{1}{P}y^T K_L^{-1} y\,.
\end{equation} 
In particular, we expect the finite-width network to outperform its infinite-width counterpart whenever $s_L < 1$. In panel (a) and (c) of Fig. \ref{fig:2} we check this prediction for deep architectures with ReLU activation on the same regression tasks employed in the 1HL case. Notice that $s_L$ quickly diverges to infinity as the number of hidden layers $L$ grows. The reason for this is simply that the ReLU NNGP kernel $K_L$ develops at least one zero eigenvalue as $L \to \infty$. This ultimately occurs because each element of the matrix $K_L$ converges to zero as $L$ grows (see panel (b) and (c) of Fig. \ref{fig:2}), as one can easily check by looking at the explicit recurrence relation for the NNGP ReLU kernel. \cite{cho2009kernel, LeeGaussian}. We note that this singularity can be equivalently thought as the fixed point of the discrete dynamical map defined by the recurrence relation for the NNGP kernel and therefore it might be worth investigating the relation between the generalisation performance in our asymptotic limit and the line of work on the edge of chaos in random neural networks \cite{ganguli2016chaos, yang2017residual}. 

Equation~\eqref{correction} provides an additional link with Student's $t$ inference. In fact, the same criterion has been found by Tracey and Wolpert \cite{tracey2018student} in the study of Bayesian optimization with Student's $t$-processes. Here the authors show that the value of $s_L$ determines whether the Student's $t$-process they consider has a larger/smaller variance than the corresponding GP with the same kernel.

\begin{figure}[t]
\includegraphics[width=0.45\textwidth]{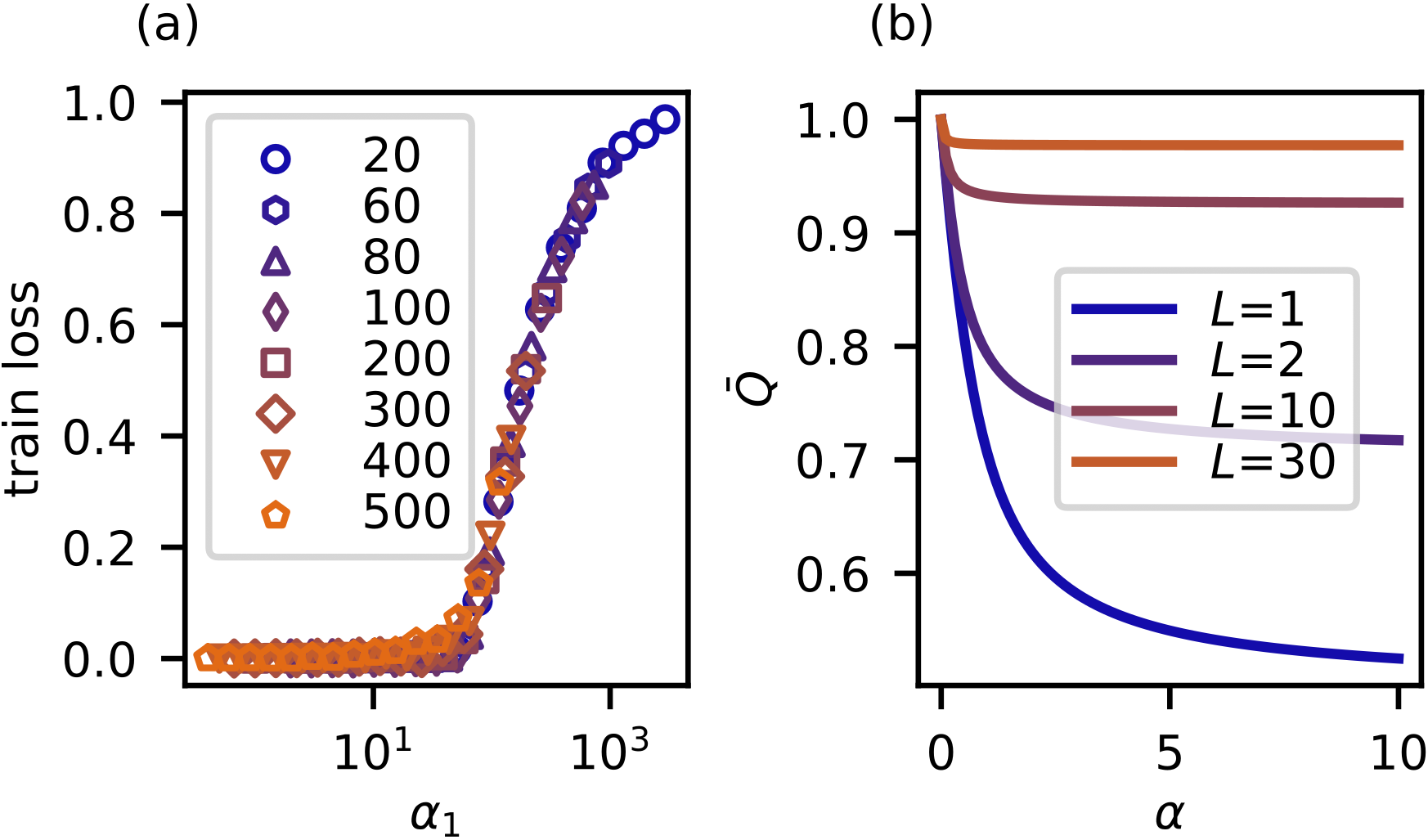}
\caption{(Left panel) Training loss of different one-hidden layer architectures trained on a completely random task (i.e. both the inputs $\mathbf x \in \mathbb{R}^{N_0}$ with $N_0 = 50$ and the scalar outputs $y$ are i.i.d. random variables sampled from a normal distribution with zero mean and unit variance) as a function of $\alpha_1$. At the moment we can not capture this universal phenomenon with our theory, which only describes the overparametrised limit where the training error is exactly zero. (Right panel) The numerical evaluation of the solution $\bar Q$ is shown in the case of ReLU activation function and isotropic network $\alpha_{\ell} =\alpha$ $\forall \ell$, for different depths $L$. As $L$ grows ($L \sim 30$), the parameter $\bar Q$ quickly approaches $1$ for all $\alpha$, suggesting that also DNNs in the asymptotic regime converge to a kernel limit in the sequential limit where the depth $L$ is taken to infinity after $P,N$.}
\label{fig:3}
\end{figure}

\section{Discussion}

\label{sec:discussion}

In our work we have described a strategy to investigate the statistical mechanics of deep neural networks beyond the infinite-width limit, that is in the finite asymptotic regime $P, N_\ell \to \infty$ at $\alpha_\ell = P/N_\ell >0$ as opposed to the infinite-width $\alpha_\ell =0$. In the 1HL case, we conjecture that our evaluation is exact in the above thermodynamic limit. As such, we do not expect any additional corrections to the result, at least in the asymptotic regime. In particular, we have found a closed expression for the generalisation error that in principle provides a Bayesian estimator of the generalisation capabilities of fully-connected architectures for any given empirical dataset, provided that the chosen architecture is capable of perfectly fitting the trainset.

For the case of finite depth $L>1$ networks, it should be possible, at least in principle, to take systematically into account non-Gaussian corrections to the saddle-point action to check whether these are relevant or not for the theory at finite width, since the assumptions we made in deriving the results are clear {\cite{roberts2022book}} (see also Methods).

From the mathematical perspective, we find the link with Student's $t$-processes very promising. The precise characterization of this mapping and its limits of validity represent a research line for future investigation.

Notably, our theory predicts that a kernel limit should also appear in the asymptotic regime as the depth $L$ approaches infinity. This could be checked, for instance, considering the isotropic limit $\alpha_{\ell} = \alpha$ $\forall \ell$ and ReLU activation. Here one can numerically solve the saddle-point equation for $\bar Q$ at large $L$ and verify that $\bar Q \to 1$ for all $\alpha$, as shown in panel (b) Fig.~\ref{fig:3}. As such, also in this limit we expect an equivalence with a kernel theory with kernel given by $K_{\infty} (C)$. Note that from our framework it is clear that we are taking the depth $L$ infinity only after $P,N$. As such we are not making claims about the challenging simultaneous limit $L,N\to \infty$ at fixed $L/N$, as done for instance in Refs. \cite{hanin2023bayesian, hanin2023random,yaida2020nonGauss,roberts2022book}.

It is fair to stress that our theory only describes the equilibrium regime of zero train loss, so that our analysis should not apply in the regime $P/N_1 \gg 1$.
Interestingly, numerical simulations performed with 1HL architectures of varying width and random training labels show that the train loss follows a universal behavior \cite{gerace2022gaussian, zavatone2022contrasting} w.r.t. $\alpha_1$ (see Fig. \ref{fig:3}, panel (a)) also in the regime where the DNN is not capable to perfectly fit the data. It would be desirable to develop a theory that also captures this phase.

Another interesting aspect to understand is the degree to which this mean-field static analysis can be extended beyond equilibrium in order to assess the full training dynamics; such a theory would indeed make it possible to investigate the performance of the many (often heuristics) learning algorithms employed to train deep neural networks. 

We conclude by pointing out that it would be interesting to compare our theory at fixed data with the data-averaged cases studied in \cite{zavatone2022contrasting, cui2023optimal} and to extend our results to convolutional layers, as done in the infinite-width case in Ref. \cite{novak2019bayesian}.

\section*{Methods}
\subsection{Setting of the learning problem and notation}
\label{sec:method_setting}

We consider deep neural networks $f_{\textrm{DNN}} (\mathbf x)$ with $L$ fully-connected hidden layers, where the pre-activations of each layer $h_{i_{\ell}}^{(\ell)}$ ($i_{\ell} = 1,\dots, N_{\ell}$; $\ell = 1, \dots, L$) are given recursively as a non-linear function of the pre-activations at the previous layer $h_{i_{\ell-1}}^{(\ell-1)}$ ($i_{\ell-1}= 1, \dots, N_{\ell-1}$):
\begin{align}
h_{i_\ell}^{(\ell)} &= \frac{1}{\sqrt {N_{\ell-1}}} \sum_{i_{\ell-1}=1}^{N_{\ell-1}} W^{(\ell)}_{i_{\ell}i_{\ell-1}} \sigma\!\left(h_{i_{\ell-1}}^{(\ell-1)}\right) + b_{i_\ell}^{(\ell)}\,,\\
h_{i_1}^{(1)} &= \frac{1}{\sqrt {N_{0}}} \sum_{i_0=1}^{N_{0}} W^{(1)}_{i_1 i_0} x_{i_0} + b_{i_1}^{(1)}\,
\end{align}
where $W^{(\ell)}$ and $b^{(\ell)}$ are respectively the weights and the biases of the $\ell$-th layer, whereas the input layer has dimension $N_0$ (the input data dimension). $\sigma$ is a non-linear activation function and it is common to each layer. We add one last readout layer and we define the function implemented by the deep neural network as:
\begin{equation}
\label{eq:f_DNN}
f_{\textrm{DNN}} (\mathbf x) = \frac{1}{\sqrt{N_L}} \sum_{i_L=1}^{N_L} v_{i_L} \sigma \!\left[ h_{i_L}^{(L)} (\mathbf x)\right]\,,
\end{equation}
where $\mathbf v$ is the vector of weights of the last layer. 

The average training error at a given inverse temperature $\beta$ is given by:
\begin{equation}
\braket{\epsilon_{\textrm t}} = \frac{1}{P}\int \!\mathcal D \theta\, [\mathcal L(\theta)-\mathcal L_\text{reg}(\theta)] \frac{e^{-\beta \mathcal L(\theta)} }{Z}\,,
\label{eq:err_t_def}
\end{equation}

Training and test errors (as defined in equation \eqref{eq:err_g_def}) represent two special observables, but more in general, for an arbitrary observable $O$ we have:
\begin{equation}
\braket{O} = \int \!\mathcal D \theta\, O(\theta) \frac{e^{-\beta \mathcal L(\theta)}}{Z}\,.
\end{equation}

\subsection{The Breuer-Major theorem as a justification for the Gaussian equivalence in shallow networks}

The Breuer-Major theorem and its extensions deal with the following sequence of random variables:
\begin{equation}
S_N = \frac{1}{\sqrt N} \sum_{i=1}^N c_i F(x_i) \quad N\geq 1\,.
\end{equation}
Clearly, if the distribution of the vector $\mathbf x = (x_1,\dots, x_N)$ is factorized over its coordinates, i.e. $p(\mathbf x) = \prod_i p(x_i)$ and $F (x) = x$, the random variable $S = \lim_{N\rightarrow \infty} S_N$ is normal distributed as long as the mean $ \mathbb{E}(x_i) = 0$, the variance $\mathbb{E}( x_i^2 )$ is finite and the $c_i$'s satisfy the so-called Lindeberg's condition. This is also true whenever $F$ is a well-behaved non-linearity.

The Breuer-Major theorem essentially extends this result to generic GPs, providing sufficient conditions on the covariance matrix of the GP and on the non-linearity $F$ that guarantee convergence of $S_N$ to the normal distribution. 
We report here the modern statement of the theorem given in Ref. \cite{NourdinQuantitative}.

We first consider a stationary (unidimensional) GP $x = (x_k)_{k \in \mathbb Z}$. Stationarity  --which is not essential and will be replaced by a weaker condition in the following-- amounts to require that the covariance of the process $C_{ij} = \mathbb E (x_i x_j)$ is a function of the difference $i-j$, i.e. $C_{ij} = C(i-j)$. The only technical condition to be imposed on the non-linear function $F$ is to have well-defined \emph{Hermite rank} $R$. The Hermite rank is the smallest positive integer that appears in the decomposition of $F$ over the Hermite polynomials:
\begin{equation}
F(x) = \sum_{k=R}^{\infty} f_k \He_k(x)\,,
\end{equation}
where $\He_k(x)$ is the $k$-th Hermite polynomial and $f_k$ the coefficient of the expansion. For many reasonable activation functions $F$, $R=1$.  

\begin{theorem}[Breuer and Major, 1983]
Let  $x = (x_k)_{k \in \mathbb Z}$ be a stationary unidimensional GP with covariance $C(i-j)$. Let $\mathbb E\left[ F(x_1)\right] = 0$ and $\mathbb E \left[F^2(x_1)\right] < \infty$ and assume that the function $F$ has Hermite rank $R \geq 1$. Suppose that:
\begin{equation}
\sum_{j\in \mathbb Z} \lvert C_{1j} \rvert^R  < \infty\,.
\end{equation}
Then $\sigma^2 := \mathbb E\left[ F(x_1)^2\right] + 2 \sum_{j =1}^\infty  \mathbb E\left[ F(x_1) F(x_j)\right]$ is finite. Moreover, one has that the sequence of random variables
\begin{equation}
S_N = \frac{1}{\sqrt N} \sum_{i=1}^N F (x_i)\, \quad N \geq 1
\label{vanillaBM}
\end{equation}
converges in distribution to $\mathcal N(0,\sigma^2)$, i.e. to a Gaussian distribution with zero mean and variance $\sigma^2$.
\end{theorem}

For our scopes we will need a slightly stronger statement than the one just mentioned: (i) in our calculation the covariance will not be stationary and (ii) we will need to consider a more general sequence of nonlinear functions $c_i F (x_i)$, such that each term of the sum (\ref{vanillaBM}) is weighted by a factor $c_i \neq 1$.

It has been shown, already in the original reference \cite{BM}, that the hypothesis of stationarity can be weakened and replaced with a requirement of uniform convergence of the elements of the covariance, namely:
\begin{equation}
\sum_{j \in \mathbb Z} |C_{ij}|^R < B_0 \quad \forall i \in \mathbb Z\,,
\label{eq:BMnonstationary}
\end{equation}
where $B_0$ is a positive finite constant. Extensions (i) and (ii) have been addressed more recently by Bardet and Surgailis in \cite{bardet2013}, as we report in the following. Let $\mathbf{x}^N$ be an $N$-dimensional Gaussian vector, such that $\expect[x^N_i] = 0$, $\expect[(x^N_i)^2] = 1$. Now define $C^N_{ij} = \expect[x^N_i x^N_j]$. For a given integer $m\ge 1$, assume
\begin{align}
   & \sup_{N \ge 1} \max_{1\le j \le N} \sum_{i=1}^N |C^N_{ij}|^m < \infty\,, \label{eq:bardetMax}\\
   & \sup_{N \ge 1} \frac{1}{N}  \sum_{\substack{1\le i, j \le N\\ |i - j| > K}} |C^N_{ij}|^m \underset{K\to\infty}{ \longrightarrow } 0 \,. \label{eq:bardetSum}
\end{align}
Take also $\mathbb{L}^2_0(x) = \left\{ f : \expect f(x) = 0,   \expect f^2(x)  < \infty \right\}$, where $x$ is a standard normal variable. Then
\begin{theorem}[Bardet and Surgailis~\cite{bardet2013}, 1.ii]
Assume~\eqref{eq:bardetMax},~\eqref{eq:bardetSum}. Let $f^N_i \in \mathbb{L}^2_0 (x)$ ($N\ge 1$, $1\le i \le N$) be a sequence of functions all having Hermite rank $m$ at least one. Assume that there exist a $\mathbb{L}^2_0(x)$-valued continuous function $\phi_\tau$, $\tau \in [0,1]$, such that
\begin{equation}
    \sup_{\tau \in (0,1]} \expect [f^N_{[\tau N]}(x) - \phi_\tau(x)]^2 \underset{N\to\infty}{\longrightarrow} 0\,. \label{eq:bardetPhi}
\end{equation}
Moreover, let
\begin{equation}
    (\sigma^N)^2 = \expect \left[\frac{1}{\sqrt{N}} \sum_{i = 1}^N f^N_i(x^N_i) \right]^2  \underset{N\to\infty}{\longrightarrow} \sigma^2 \,,
\end{equation}
where $\sigma^2 >0$. Then
\begin{equation}
    \frac{1}{\sqrt{N}} \sum_{i = 1}^N f^N_{i}(x^N_i) \underset{N\to\infty}{\overset{d}{\longrightarrow}} \mathcal{N}(0,\sigma^2)\,,
\end{equation}
where $\underset{N\to\infty}{\overset{d}{\longrightarrow}}$ denotes convergence in distribution.
\end{theorem}
The hypotheses of this theorem should be taken as conditions on the activation function $\sigma$, on the rescaled input covariance matrix $C_{\mu\nu}$ and on the dominant configurations $\bar{s}$ in the Fourier integral~\eqref{eq:P(s|data)} in order to justify our Gaussian ansatz (see below, Eq.~\eqref{eq:q_is_gaussian}).

\subsection{Sketch of the calculation of the effective action in the Bayesian setup for one-hidden layer fully-connected neural networks}

\label{sec:Zcalc}

We now discuss the salient aspects of the calculation. The starting point is the following partition function:
\begin{multline}
Z = \int\! \prod_{i_1}^{N_1} \!\diff v_{i_1}\!\prod_{i_1,i_0}^{N_1,N_0}\! \diff w_{i_1 i_0}\, \exp\Biggl\{-\frac{\lambda_1}{2} \sum_{i_1}^{N_1} v_{i_1}^2 -\frac{\lambda_0}{2} \lVert w \rVert^2 \\
-\frac{\beta}{2} \sum_{\mu}^P \left[ y^\mu -\frac{1}{\sqrt{N_1}} \sum_{i_1}^{N_1}  v_{i_1} \sigma\!\left( \sum_{i_0}^{N_0}\frac{w_{i_1,i_0}x_{i_0}^\mu}{\sqrt{N_0}} \right) \right]^2\Biggr\}\,.
\end{multline}
where $w = W^{(1)}$ and we took $b^{(1)}=0$ without loss generality \footnote{One can map a system with non-zero biases in a zero-bias one increasing by one the dimensions of the input and of the activations at each layer. The original biases are then trivially mapped in the extra weights of the augmented system.}.
The first step is to decouple the weights of the different layers in the loss function. This can be done including standard identities built over two families of Dirac deltas, one for the pre-activations of the hidden layer and one for the output of the network:
\begin{align}
&1=\int \prod_\mu^P ds^\mu \delta\!\left[s^\mu - \frac{1}{\sqrt{N_1}} \sum_i^{N_1} v_{i_1} \sigma(h_{i_1}^\mu)\right] \,, \label{eq:delta_s}\\
&1=\int \prod_\mu^P \prod_{i_1}^{N_1}  dh_{i_1}^\mu \delta\!\left(h_{i_1}^\mu - \frac{1}{\sqrt{N_0}} \sum_{i_0}^{N_0} w_{i_1 i_0} x_{i_0}^\mu \right) \,.
\label{eq:delta_h}
\end{align}

By using a standard Fourier representation of these deltas, which introduces the conjugate variables $\bar{h}^\mu_{i_1}$ and $\bar{s}^\mu$, we can perform the gaussian integrals on the internal and external weights: 
\begin{equation}
    \begin{split}
&        Z={} \int \prod_\mu^P \frac{\diff s^\mu \diff \bar{s}^\mu}{2\pi} e^{-\frac{\beta}{2}\sum_\mu\left( y^\mu - s^\mu \right)^2 + i\sum_\mu^P s^\mu \Bar{s}^\mu} \\
        &\times \Biggl\{\int \prod_\mu^P \frac{\diff h^\mu \diff \Bar{h}^\mu}{2\pi} e^{i\sum_\mu^P h^\mu \Bar{h}^\mu -\frac{1}{2 \lambda_1 N_1} \left[ \sum_\mu^P \Bar{s}^\mu \sigma\left(h^\mu\right) \right]^2}\\
        & \hspace{8em}\times e^{-\frac{1}{2 \lambda_0 N_0 }\sum_{i_0}^{N_0}\left(\sum_\mu^P \Bar{h}^\mu x^\mu_{i_0}\right)^2}\Biggr\}^{N_1}\,,
    \end{split}
    \label{eq:Z_intermediate}
\end{equation}
where we used the fact that the integrals on $h_{i_1}^\mu$ and $\bar{h}_{i_1}^\mu$ can be factorized on the index $i_1$. The integral over the $\bar{h}^\mu$ is Gaussian and can be solved:
\begin{equation}
\int \prod_\mu^P \frac{\diff \Bar{h}^\mu}{2\pi} e^{i\sum_\mu^P h^\mu \Bar{h}^\mu-\frac{1}{2 \lambda_0 N_0 }\sum_{i_0}^{N_0}\left(\sum_\mu^P \Bar{h}^\mu x^\mu_{i_0} \right)^2}  = P_1(\{h^\mu\})\,,
\end{equation}
where 
\begin{equation}
P_1(\{h^\mu\})
=\frac{e^{-\frac{1}{2}\sum_{\mu,\nu}^P h^\mu C_{\mu,\nu}^{-1} h^{\mu}  }}{\sqrt{(2\pi)^P \det C}} \,, \,\, C_{\mu\nu} = \frac{1}{\lambda_0 N_0} \sum_{i_0}^{N_0} x^\mu_{i_0} x^\nu_{i_0}\,.
\label{eq:P1}
\end{equation}
This last step requires the covariance matrix $C$ to be invertible. Note that this is false as soon as $P > N_0$, but adding a small diagonal term to $C$ solves the issue. One can explicitly check that the final result does not depend on this extra regularization.

To deal with the integral over $h^\mu$ we can include a further Dirac delta identity for the random variable $q = 1/\sqrt{\lambda_1 N_1}\sum_\mu \Bar{s}^\mu \sigma\left(h^\mu\right)$. This leaves us with the problem of finding the probability density $P(q)$. In the limit defined in \eqref{eq:finite_scaling}, this is exactly the same setting of the Breuer-Major theorems~\cite{BM,NourdinQuantitative, bardet2013}. As such, it is sufficient that both the (regularized) covariance $C$ and the activation function $\sigma$ satisfy the hypotheses of the 
theorem to guarantee that the probability distribution $P(q)$ converges in distribution to a Gaussian:  
\begin{equation}
    \begin{split}
      &P(q)  = \int  \diff^P h \, P_1(\{h^\mu\})  \delta\Biggl[q-\frac{1}{\sqrt{\lambda_1 N_1}}\sum_\mu \Bar{s}^\mu \sigma\left(h^\mu\right) \Biggr]\,,\\
      &P(q) \to \mathcal{N}_q(0,Q)\,.
    \end{split}
    \label{eq:q_is_gaussian}
\end{equation}
with variance
\begin{equation}
    \begin{aligned}
        Q(\bar{s},C) &= \frac{1}{\lambda_1 N_1} \sum_{\mu,\nu}^P \Bar{s}^\mu\left[\int \diff^P h\, P_1(\{h^\rho\}) \sigma(h^\mu)\sigma(h^\nu)\right] \Bar{s}^\nu \\
        &=\frac{1}{\lambda_1 N_1}\sum_{\mu,\nu}^P \Bar{s}^\mu K_{\mu \nu}(C)\Bar{s}^\nu\,.
    \end{aligned}
    \label{eq:K}
\end{equation}

One can show that there exist special configurations $\bar s$ in the domain of integration for which we are not allowed to invoke a Gaussian equivalence (see for instance our discussion at the end of Sec.~\ref{supp:sec:special_cases} in the supplemental material \cite{supplemental}). In our derivation, we are assuming that the contribution of these special configurations to the effective action is negligible in the thermodynamic limit.
Here we have also assumed that the variable $q$ has zero mean, a condition verified as long as
\begin{equation}
\int  \diff^P h \, P_1(\{h^\mu\})  \sigma(h^\nu)  = 0\,,
\label{methods:eq:zeromean}
\end{equation}
that is whenever $\sigma$ is zero-mean; for a more general derivation, relevant for finite-mean activation functions such as ReLU, see the supplemental material \cite{supplemental}, Sec.~\ref{supp:sec:finitemeanactivation}.

Each element of the kernel matrix $K_{\mu\nu} (C)$ can be easily reduced from a $P$-dimensional integral to a simpler two-dimensional one: 
\begin{align}
K_{\mu\nu}(C) = \int \frac{\diff t_1 \diff t_2}{\sqrt{(2\pi)^2 \det \tilde C}} e^{-\frac{1}{2} \mathbf t^T \tilde C^{-1} \mathbf t}  \sigma(t_1) \sigma (t_2)\,, 
\label{methods:eq:kernel}
\end{align}
where ${\bf t}=(t_1,t_2)^T$ and
\begin{equation}
\tilde C = \begin{pmatrix} C_{\mu\mu} & C_{\mu\nu}\\ C_{\mu\nu} & C_{\nu\nu}\end{pmatrix}\,. 
\end{equation}
is the reduced $2\times2$ input covariance matrix.
It is worth pointing out that the kernel we find here is the so-called \emph{neural network Gaussian process} (NNGP) kernel. It differs from the neural tangent kernel (NTK) that is found in the infinite-width limit of networks trained under gradient descent~\cite{NEURIPS2020_ad086f59}. The fact that the infinite-width limit of a Bayesian neural network differs from the one obtained from gradient descent is indeed known and discussed in literature~\cite{lee2019wide}. 

Now we can integrate over the variable $q$ and obtain:
\begin{equation}
    \left[\int \frac{\diff q \,e^{ -\frac{q^2}{2}-\frac{q^2}{2 Q(\bar{s},C)} }}{\sqrt{2\pi Q(\bar{s},C)}} \right]^{\frac{N_1}{2}} = \left[Q(\bar{s},C)+1\right]^{-\frac{N_1}{2}}\,.
    \label{eq:P(Q)}
\end{equation}

In the general case of finite $\alpha_1 = P/N_1$, we are only left with the integrals in $s^\mu$ and $\bar{s}^\mu$. To solve them it is convenient to introduce one final Dirac delta identity:
  \begin{equation}
  1=\int dQ \,\delta\Biggl[Q-\frac{1}{\lambda_1 N_1}\sum_{\mu,\nu} \bar{s}^\mu K(C)_{\mu\nu} \bar{s}^\nu\Biggr]\,,
\label{EQ:Q}
 \end{equation}
 where $Q \geq -1$ is now an integration variable and not a function of $\bar{s}$, so that we have removed the explicit dependence on $\sqrt{Q(\bar{s},C)+1}$ in the partition function. Finally, the integrals in $s^\mu$ and $\bar{s}^\mu$ are Gaussian once another integral representation of the delta via a conjugate variable $\bar Q$ is inserted. This allows us to get the final effective action obtained in equation \eqref{effS}.

\subsection{Exact solution of the saddle-point equations in the zero temperature limit}

The saddle-point equations obtained from \eqref{effS} considerably simplify in the zero temperature limit ($\beta \to \infty$). In particular, using the fact that the kernel $K$ has only positive eigenvalues (in the asymptotic regime $\alpha_1$, $\alpha_0$ finite), we get:

\begin{align}
\bar Q = \frac{1}{1+Q}\,, \qquad Q = +\frac{\alpha_1}{\bar Q} -\frac{\alpha_1}{\bar Q^2 }\frac{1}{P} y^T\left(\frac{K}{\lambda_1}\right)^{-1}y\,.
\end{align} 
Given the condition $Q\geq-1$, the unique exact solution for $\bar Q$ is positive and reads:
\begin{equation}
\bar Q^* = \frac{ \sqrt{(\alpha_1-1)^2 + 4 \alpha_1 \frac{1}{P} y^T\left(\frac{K}{\lambda_1}\right)^{-1}y}-(\alpha_1 -1)}{2}\,.
\end{equation}

\subsection{Predictors statistics}
%

The main observable we are interested in is the generalisation error~\eqref{eq:err_g_def}. We can proceed along the same lines of the calculation performed in Sec.~\ref{sec:Zcalc} introducing, other than the variables $s^\mu$, $h_i^\mu$ defined by~\eqref{eq:delta_s}, \eqref{eq:delta_h}, additional variables $s^0$, $h_i^0$ that describe output and pre-activations of the new test example. 
We thus get:
\begin{equation}
\begin{split}
&\braket{\epsilon_\text{g}(\mathbf{x}^0,y^0)} =\frac{1}{Z} \int \frac{\diff s^0 \diff \bar{s}^0}{2\pi} \int \prod_{\mu=1}^P \frac{\diff s^\mu \diff \bar{s}^\mu}{2\pi} \, (y^0-s^0)^2  \\
&\quad \times e^{-\frac{\beta}{2}\sum_{\mu=1}^P (y^\mu-s^\mu)^2 + i\sum_{\mu=1}^P s^\mu \bar{s}^\mu +is^0\bar{s}^0  }\\
&\quad \times\Biggr[1+\frac{1}{\lambda_1 N_1}\Biggl(\sum_{\mu,\nu=1}^P \bar{s}^\mu K_{\mu\nu} \bar{s}^\nu \\
& \hspace{5em}+ 2 \bar{s}^0\sum_{\mu=1}^P \bar{s}^\mu \kappa_{\mu}(\mathbf x^0)  + (\bar{s}^0)^2 \kappa_{0}(\mathbf x^0) \Biggr)\Biggr]^{-\frac{N_1}{2}}\,,
\end{split}
\label{eq:gen_err0}
\end{equation}
where $\kappa_\mu$ and $\kappa_0$ are respectively the train-test and the test-test kernel
integrals defined as in~\eqref{eq:K} when the covariance matrix involves the test input, namely:
\begin{align}
&\kappa_\mu = \int \frac{dt_1 dt_2}{\sqrt{(2\pi)^2 \det \tilde C_\mu}} e^{-\frac{1}{2} \mathbf t^T \tilde C_\mu^{-1} \mathbf t}  \sigma(t_1) \sigma (t_2)\,, \label{kmu}\\
&\kappa_0 = \int \frac{dt}{\sqrt{2\pi C_{00}}} e^{-\frac{t^2}{2 C_{00}}}  \sigma(t)^2\,, \label{k0}
\end{align}
where
\begin{align}
&\begin{gathered}
\tilde C_\mu = \begin{pmatrix} C_{\mu\mu} & C_{\mu0}\\ C_{\mu0} & C_{00}\end{pmatrix}\,, \quad C_{\mu 0} =  \frac{1}{\lambda_0 N_0} \sum_{i_0}^{N_0} x^\mu_{i_0} x^0_{i_0}\,,\\
 C_{0 0} =  \frac{1}{\lambda_0 N_0} \sum_{i_0}^{N_0} (x^0_{i_0} )^2\,.
 \end{gathered}
\end{align}
Now we can introduce the order parameters $Q$ and $\bar{Q}$ via equation (\ref{EQ:Q}) and their Fourier representation and perform the integration over all the ${s^\mu, \bar{s}^\mu}$ and over the $\bar{s}^0$. Doing so yields a single integral in $s^0$ and integrals on $Q$ and $\bar{Q}$. 
\begin{align}
&\braket{\epsilon_\text{g}(\mathbf{x}^0,y^0)} = \frac{1}{Z}  \int \frac{\diff Q \diff \bar{Q}}{2\pi} e^{-\frac{N_1}{2}S(Q,\bar{Q})} \notag\\
& \hspace{9em} \times \int \frac{\diff s^0 (y^0-s^0)^2}{\sqrt{2\pi\sigma^2}}  e^{-\frac{(s^0+\Gamma_1)^2}{2\sigma_1^2}} ,
\end{align}
with
\begin{align}
&\hphantom{^2}\Gamma_1 =\frac{\bar Q}{\lambda_1}\sum_{\mu\nu} \kappa_\mu(\mathbf x^0) \left(\frac{\mathbb{1}}{\beta}+\frac{\bar{Q}}{\lambda_1}K\right)^{-1}_{\mu\nu}\; y_\nu, \nonumber \\
&\sigma_1^2 = \frac{\bar{Q}}{\lambda_1}\Biggl[\kappa_0(\mathbf x^0)\label{eq:generc_bias_var}\\
&\hspace{4em} -\frac{\bar{Q}}{\lambda_1} \sum_{\mu\nu} \kappa_\mu(\mathbf x^0) \left(\frac{\mathbb{1}}{\beta}+\frac{\bar{Q}}{\lambda_1}K\right)^{-1}_{\mu\nu}\; \kappa_\nu (\mathbf x^0)\Biggr] \nonumber
\end{align}
We can then unfold the easy integrals in $s^0$ and evaluate the result on the saddle point solution.
The generalisation error is expressed in terms of $\Gamma_1$ and $\sigma_1^2$ as in equation \eqref{Eps_g}. Taking the $\beta \to \infty$ limit in equations \eqref{eq:generc_bias_var} yields the expressions in \eqref{Eps_g}.

\subsection{Constraints on the scaling of the size of the dataset $P$ with the input dimension $N_0$}
\label{sec:hypothesis}
In this section we address the additional constraints to the thermodynamic scaling ($P,N_1\rightarrow \infty$ with $\alpha_1 = P/N_1$ finite) that may come from the hypotheses of the Breuer-Major on the covariance matrix $C$. The only stringent condition to verify is equation~\eqref{eq:BMnonstationary}, that is
\begin{equation}
\sum_{\mu=1}^P |C_{\mu \nu}|^R < B_0 \qquad \forall \,\nu=1,\dots,P\,,
\label{eq:BM_Cmunu}
\end{equation}
where $B_0$ is a given finite constant and $R$ the Hermite rank of the activation function $\sigma$. In the case of inputs $\mathbf x$ with i.i.d. standard Gaussian coordinates, $C_{\mu\nu}$ is a Wishart random matrix with off-diagonal entries of order $1/\sqrt{N_0}$ and random signs: after taking the absolute value, the sum in Eq~\eqref{eq:BM_Cmunu} is of order $P (N_0)^{-R/2}$. Note that this provides an infinite class of activation functions (those with Hermite rank $R \geq 2$) where we can safely work at least at finite $\alpha_0 = P/N_0$. For activation functions with Hermite rank $R = 1$ (such as Erf or ReLU) we cannot provide such a guarantee by only looking at the hypothesis of the Breuer-Major theorem. It is also worth stressing that, given any odd (non-odd) activation function $\sigma (x)$ with Hermite rank $R=1$, it is easy to engineer a new  reasonable activation function with Hermite rank $R=3$ ($R=2$), just by replacing the old activation function with a new one $\sigma_1 (x) = \sigma (x) - g_1 x$, where the coefficient $g_1 = \langle \sigma (x) \He_1 (x) \rangle$ and the average is over a normal distribution of zero mean and unit variance. 

We observe that there is at least one case of activation function with $R=1$ where the derivation goes through at finite $\alpha_0$, i.e. the linear function $\sigma(x) = x$ (in this case we can obtain the result at finite $P, N_1, N_0$, as done also in Ref. \cite{hanin2023bayesian}). 
In the supplemental material \cite{supplemental}, Sec.~\ref{supp:sec:special_cases}, we examine the specific case of quadratic activation $\sigma(x) = x + x^2$ (that has $R=1$), deriving the final effective action without employing the BM theorem. As in the linear case, this derivation goes through at finite $\alpha_0$.
We are thus led to think that the scaling $P= O(\sqrt{N_0})$ suggested for $R=1$ is overly-pessimistic.

\subsection{Generalisation to deep neural networks with a finite number of hidden layers $L > 1$ and zero-mean activation}
In the same spirit of the 1HL calculation, we introduce $L$ sets of auxiliary variables $h_{ i_\ell}^{\mu}$ (where $i_\ell=1,\dots, N_\ell$) that are equal to the pre-activations at each layer. The strategy to perform the calculation is to show that the probability distribution of the preactivations at each layer $P_{\ell}(\{h_{i_\ell}^\mu\})$ can be computed recursively, starting from the input layer. We notice that this is conceptually different from the backpropagating kernel renormalisation group introduced in Ref. \cite{SompolinskyLinear}. It is still a kernel renormalisation group, but forward-propagating, and represents a generalisation to NNTPs of the kernel recurrence arising in NNGPs~\cite{LeeGaussian}. In practice, our approach amounts to a systematic, layer-by-layer description of the pre-activation statistics by the Student's t distribution that we have shown to appear in the 1HL case. This can be seen as a quantitative correction to the standard Gaussian statistics that is recovered in the infinite width limit. At the moment we are not able to re-derive the same result using the backpropagating method introduced in~\cite{SompolinskyLinear}.

Let us start by integrating the weights of the first layer. This defines a probability distribution over the pre-activations of the first layer via:
\begin{equation}
\begin{aligned}
P_1(\{h_{i_1}^\mu\}) &= \int \mathcal D W^{(1)} \prod_{i_1, \mu} \delta\left(h_{i_1}^{\mu} -\frac{1}{\sqrt {N_{0}}} \sum_{i_0=1}^{N_{0}} W^{(1)}_{i_1 i_0} x_{i_0}^\mu  \right)\\
&= \prod_{i_1=1}^{N_1} \frac{e^{-\frac{1}{2} \sum_{\mu \nu} h_{i_1}^{\mu} C^{-1}_{\mu\nu}  h_{i_1}^{\nu} }}{\sqrt{(2\pi)^P \det C}}\,.
\end{aligned}
\end{equation}
where $C$ is defined in~\eqref{eq:P1}. This result is straightforward and it is valid for any $N_0$, $P$ and $N_1$, since the prior for the weights is gaussian. At the second layer we have:
\begin{align}
P_2(\{h_{i_2}^\mu\}) 
&= \int \mathcal D W^{(2)} \mathcal D h_1 P_1 (\{h_{i_1}^{\mu}\}) \notag\\
&\times \prod_{i_2, \mu} \delta\left(h_{i_2}^{\mu} -\frac{1}{\sqrt {N_{1}}} \sum_{i_1=1}^{N_{1}} W^{(2)}_{i_2 i_1} \sigma(h_{i_1}^\mu)  \right) 
\end{align}
We now introduce conjugate variables $\bar h^{\mu}_{i_2}$ to the activation of the second layer and the calculation proceeds as in the case of 1HL architectures. To make analytical progress we need to make two fundamental approximations: (i) assuming that the set of random variables $q_{i_2} = 1/(\sqrt{N_1} \lambda_1) \sum_\mu \bar h^{\mu}_{i_2} \sigma(h^\mu)$, where $\mathbf h \sim \mathcal N(0, C)$, is Gaussian-distributed; (ii) neglecting correlations between different pre-activations of the second hidden layer. In conclusion we get:
\begin{multline}
P_2(\{h_{i_2}^\mu\}) = \int \diff Q_1 \diff\bar Q_1 e^{-\frac{N_1}{2} (-Q_1 \bar Q_1 + \log (1+Q_1))}\\
\times \prod_{i_2=1}^{N_2} \frac{e^{-\frac{1}{2} \sum_{\mu\nu}h_{i_2}^{\mu}\left(\bar Q_1 K(C)/\lambda_1\right)^{-1}_{\mu \nu} h_{i_2}^{\nu}}}{\sqrt{(2\pi)^P \det (\bar Q_1  K(C)/\lambda_1)}}\,,
\label{preact12}
\end{multline}
where $K(C)$ is defined by equation \eqref{eq:K}. Notice that except for the integration over the two variables $Q_1$ and $\bar Q_1$, this is the same as the probability distribution of the 1HL system \eqref{eq:P1} if we replace $C$ with $\bar Q_1  K(C)/\lambda_1$. This reasoning can be easily iterated across layers and gives:
\begin{multline}
P_L (\{h_{i_L}^\mu\}) = \!\!\int \prod_{\ell =1}^{L-1} \!\diff Q_\ell \diff \bar Q_\ell e^{-\sum_{\ell=1}^{L-1} \frac{N_\ell}{2}\left[ -Q_\ell \bar Q_{\ell} + \log(1+Q_\ell)\right]}\\
\times \prod_{i_L=1}^{N_L} \frac{e^{-\frac{1}{2} \sum_{\mu\nu}h_{i_L}^{\mu}\left( K^{(R)}_{L-1}(\{\bar Q_\ell\} )\right)^{-1}_{\mu \nu} h_{i_L}^{\nu}}}{\sqrt{(2\pi)^P \det (K^{(R)}_{L-1}(\{\bar Q_\ell\}))}}\,,
\end{multline}
where $K^{(R)}_{\ell}(\{\bar Q_\ell\})$ is a renormalised kernel that satisfies the recurrence relation in equation \eqref{K_LQ}.

The computation of the generalisation error over a new example $(\mathbf x^0, y^0)$ gives:
\begin{equation}
\begin{split}
\braket{\epsilon_\text{g}(\mathbf{x}^0,y^0)} &= (y^0-\Gamma_L)^2+\sigma^2_L
\end{split}
\label{methods:eq:gen_err0deep}
\end{equation}
where $\Gamma_L$ and $\sigma^2_L$ are defined respectively in Eqs. \eqref{GL} and \eqref{sigmaL}. Note that $\kappa^{(R)}_{L\mu}$, $\kappa^{(R)}_{L0}$ are recursive kernels that generalise the train-test and test-test kernels (\ref{kmu})-(\ref{k0}). They are defined starting from equation \eqref{K_LQ} where the kernel $K$ is now evaluated with the covariance matrix $C$ involving train-test or test-test points.
Note that $L$-hidden layers generalisation error is found replacing the 1HL kernel with its recursive generalisation (\ref{K_LQ}).

\subsection{Numerical experiments}
\label{sec:experiments}

\subsubsection{Network architectures}
We perform numerical experiments with deep fully-connected architectures trained on two regression tasks in computer vision. In particular we use the 0 and 1 classes of the MNIST and CIFAR10 datasets, which for the latter correspond to the labels ``cars'' and ``planes''. Examples from CIFAR10 are coarse grained to $N_0=28 \times 28$ pixels and converted to grayscale. 

To test our theory in the zero-mean activation function case, we used the $\text{Erf}$ function, for which the NNGP kernel can be computed analytically~\cite{PANG2019270}: 
\begin{equation}
K^{\textrm{Erf}}_{\mu\nu} (C) = \frac{2}{\pi} \arcsin\left(\frac{2 C_{\mu\nu}}{\sqrt{\left(1+ 2 C_{\mu\mu}\right)\left(1+ 2 C_{\nu\nu}\right)}}\right)\,.
\end{equation}
In Fig. \ref{fig:2} we train networks with $\sigma = \text{ReLU}$. The kernel can be computed analytically also in this case \cite{cho2009kernel} and reads: 
\begin{align}
K^{\textrm{ReLU}}_{\mu\nu} (C) &= \sqrt{C_{\mu \mu} C_{\nu \nu}} \, \kappa \left( \frac{C_{\mu\nu}}{\sqrt{C_{\mu \mu} C_{\nu \nu}}} \right)\,, \\ \nonumber  
\kappa(x) &= \frac{1}{2\pi} \left[x(\pi - \arccos(x))+ \sqrt{1-x^2}\right]\,.
\end{align}

\subsubsection{Sampling from the Bayesian posterior}

To ensure convergence of the posterior weights distribution to the Gibbs ensemble, we train our networks using a discretised Langevin dynamics, similarly to what is done in \cite{SompolinskyLinear,seroussi2023natcomm}. At each training step $t$ the parameters $\theta = \lbrace W^\ell, v\rbrace $ are updated according to: 
\begin{equation}
    \theta(t+1) =  \theta(t) - \eta \nabla_\theta \mathcal{L}(\theta(t)) +\sqrt{2T\eta}\epsilon(t)
\end{equation}
where $T=1/\beta$ is the temperature, $\eta$ is the learning rate, $\epsilon(t)$ is a white Gaussian noise vector with entries drawn from a standard normal distribution, and the loss is the one defined in equation \eqref{loss}. We employ $T = \eta =  10^{-3}$ throughout all the experiments. This is sufficient to approximate the $T=0$ dynamics in the regime we are considering. This dynamics requires $10^5$/$10^6$ steps to reach thermalisation, depending on the sizes of the dataset and network. We extract the generalisation loss within a single run: after the train error has reached its minimum and the test loss is thermalised, we average test loss values every $10^3/10^4$ epochs (depending again on the magnitude of $P$, $N_\ell$). For the sake of completeness, we report the best test accuracy achieved on both datasets by 1HL architectures: $0.86$ on CIFAR10 with $P = 3000$ and $\lambda_1 = 1000$, $0.999 $ on MNIST with the same Gaussian prior and $P = 1000$. The train accuracy is always $1$. Additional comments on the technical issues encountered in simulating the Bayesian dynamics are discussed in the Supplemental material \cite{supplemental} in Sec. \ref{supp:sec:numerical_issues}.

\section*{Data availability}
The CIFAR10 and MNIST datasets that we used for all our experiments are publicly available online, respectively at \url{https://www.cs.toronto.edu/~kriz/cifar.html} and \url{http://yann.lecun.com/exdb/mnist/}. 

\section*{Code availability}
The code used to perform experiments, compute theory predictions and analyze data is available at: \url{https://github.com/rpacelli/FC\_deep\_bayesian\_networks} \cite{FC_deep_bay_netwoks}.

\bibliography{biblio.bib}

\section*{Acknowledgements}
M. P. has been supported by a grant from the Simons Foundation (grant No. 454941, S. Franz). P. R. acknowledges funding from the Fellini program under the H2020-MSCA-COFUND action, Grant Agreement No. 754496, INFN (IT).
The authors would like to thank Silvio Franz, Luca Molinari, Fabian Aguirre-L\'{o}pez, Raffaella Burioni, Alessandro Vezzani, Riccardo Aiudi, Federico Bassetti, Bruno Bassetti and the Computing Sciences group at Bocconi University in Milan for discussions and suggestions.

\section*{Author contribution}
P.R., S.A. and M.P. performed the analytical calculations, supported by F.G, M.G. and R.P. Numerical experiments, data analysis and data visualization were carried out  by R.P. All the authors contributed to discussing and interpreting the results and to writing and editing the manuscript. S.A and R.P. contributed equally to the work. 

\newpage
\onecolumngrid
\setcounter{section}{0}
\begin{center}
    \large{\textbf{Supplemental material}}
\end{center}

\section{Derivation of an effective action for 1HL neural networks with multiple outputs}

In this section, we sketch the calculation to derive an effective action for 1HL neural networks with multiple outputs $\kappa > 1$. We stress that $\kappa$ is finite and the case where the number of outputs scales with the width of the hidden layer $N_1$ has been the subject of investigations in \cite{aitchison2020bigger,yang2023theory,zavatone2021depth}.
We consider the following loss function:

\begin{align}
    \mathcal L &= \frac{1}{2 \kappa} \sum_{\mu=1}^P \sum_{a=1}^{\kappa} \left[ y^\mu_{a} - (f_{\textrm{DNN}}(\mathbf x^\mu))_{a} \right]^2 + \mathcal L_{\textrm{reg}}\,,\\
        \mathcal L_{\textrm{reg}} &= \frac{\lambda_1}{2\beta} \lVert v \lVert^2 + \frac{\lambda_{0}}{2\beta} \lVert W\rVert^2\,.
\label{lossA1}
\end{align}

The partition function is defined as:
\begin{equation}
Z = \int\! \prod_{a,i_1}^{\kappa,N_1} \!\diff v_{a,i_1}\!\prod_{i_1,i_0}^{N_1,N_0}\! \diff w_{i_1 i_0}\, \exp\Biggl\{-\frac{\lambda_1}{2}  \lVert v \lVert^2  -\frac{\lambda_0}{2} \lVert w \rVert^2 
-\frac{\beta}{2 \kappa} \sum_{\mu}^P\sum_{a}^{\kappa} \left[ y^\mu_{a} -\frac{1}{\sqrt{N_1}} \sum_{i_1}^{N_1}  v_{a ,i_1} \sigma\!\left( \sum_{i_0}^{N_0}\frac{w_{i_1 ,i_0}x_{i_0}^\mu}{\sqrt{N_0}} \right) \right]^2\Biggr\}\,.
\end{equation}
We can decouple the layers in the loss through the addition of Dirac deltas, noticing that there will be one additional index $a$ for the outputs.
\begin{equation}
\begin{split}
    Z =& \int \prod_{\mu,a}^{P,\kappa} \frac{\diff s_{a}^\mu \diff \bar{s}_{a}^\mu}{(2\pi)} \; \exp\left\{-\frac{\beta}{2 \kappa} \sum_{\mu,a}^{P,\kappa} \left(y_{a}^\mu- s_{a}^\mu\right)^2+i\sum_{\mu,a}^{P,\kappa} s_{a}^\mu \bar{s}_{a}^\mu\right\}  \int \prod_{\mu,i_1}^{P,N_1} \frac{\diff h_{i_1}^\mu \diff \bar{h}_{i_1}^\mu}{(2\pi)} \exp\left\{ i\sum_{\mu,i_1}^{P,N_1} h_{i_i}^\mu \bar{h}_{i_1}^\mu \right\}\\
    &\int \prod_{a,i_1}^{\kappa,N_1} \diff v_{a,i_1} \exp\left\{-\frac{\lambda_1}{2}\lVert v \rVert^2 -i \sum_{a,\mu} \bar{s}_{a}^{\mu} \sum_{i_1} \frac{v_{a,i_1} h_{i_1}^\mu}{\sqrt{N_1}} \right\} \int \prod_{i_1,i_0}^{N_1,N_0} \diff w_{i_1,i_0} \exp\left\{-\frac{\lambda_0}{2}\lVert w \rVert^2 -i \sum_{i_1,\mu} \bar{h}_{i_1}^{\mu} \sum_{i_0} \frac{w_{i_1,i_0} x_{i_0}^\mu}{\sqrt{N_1}} \right\}\,.
\end{split}
\end{equation}
The integrals over the weights $w_{i_1,i_0}$ and $v_{a,i_1}$ are Gaussian and can be performed.
As in the single-output case we can factorize the integrals in $h_{i_1}^\mu$ and $\bar{h}_{i_1}^\mu$ over the index $i_1$:
\begin{equation}
\begin{split}
 Z=& \int \prod_{\mu,a} \frac{\diff s^\mu_{a} \diff \bar{s}^\mu_{a}}{2\pi} e^{-\frac{\beta}{2 \kappa}\sum_{\mu,a}\left( y^\mu_{a} - s^\mu_{a} \right)^2 + i\sum_{\mu,a} s_{a}^\mu \Bar{s}_{a}^\mu} \\
 &\Biggl\{\int \prod_\mu^P \frac{\diff h^\mu \diff \Bar{h}^\mu}{2\pi} e^{i\sum_\mu^P h^\mu \Bar{h}^\mu -\frac{1}{2 \lambda_1 N_1}\sum_{a}^{\kappa} \left( \sum_{\mu} \Bar{s}_{a}^\mu \sigma\left(h^\mu\right) \right)^2-\frac{1}{2 \lambda_0 N_0 }\sum_{i_0}^{N_0}\left(\sum_\mu^P \Bar{h}^\mu x^\mu_{i_0}\right)^2}\Biggr\}^{N_1}\,.
\end{split}
\end{equation}
Once the integrals over the variables $\bar{h}^{\mu}$ are performed we obtain that the $h^{\mu}$ are Gaussian-distributed with zero mean and covariance matrix $C$, in analogy with the single-output case. The critical step is to consider the joint probability distribution of the following random variables:
\begin{equation}
    q_{a} = \frac{1}{\sqrt{\lambda_1 N_1}}\sum_\mu^P \bar{s}_{a}^\mu \sigma(h^\mu).
\end{equation}
As in the single-output case, in the asymptotic proportional limit $P/N_1 \sim O(1)$ we can conjecture a Gaussian equivalence, based on the reasonable assumption that the BM theorem can be generalised to the multivariate case.
We therefore have that $P(\{q_a\}) \to \mathcal{N}(0,Q)$ where now $Q$ is the covariance matrix given by:
\begin{equation}
    \begin{aligned}
        Q(\bar{s},C)_{a,b} &= \frac{1}{\lambda_1 N_1} \sum_{\mu,\nu}^P \Bar{s}^\mu_{a}\left[\int \diff^P h\, P_1(\{h^\rho\}) \sigma(h^\mu)\sigma(h^\nu)\right] \Bar{s}^\nu_{b} =\frac{1}{\lambda_1 N_1}\sum_{\mu,\nu}^P \Bar{s}^\mu_{a} K_{\mu \nu}(C)\Bar{s}^\nu_{b}\,
    \end{aligned}
\end{equation}
and $K/\lambda_1$ is the NNGP kernel, as in the single-output case. We now integrate over the set of variables $\{q_a\}$:
\begin{equation}
    \int \prod_{a}^{\kappa} \diff q_{a} \frac{1}{\sqrt{\det(Q)}} e^{-\frac{1}{2}\sum_{a}^{\kappa} (q_{a})^2-\frac{1}{2}\sum_{a,b}^{\kappa} q_{a} Q^{-1}_{a,b} q_{b}} = \det\left(\mathbb{1}_{\kappa} + \mathbf{Q} \right)^{-\frac{1}{2}}.
\end{equation}
Differently from the single-output case, we need to introduce a $\kappa \times \kappa$ matrix order parameter $Q_{a,b}$ as:
\begin{equation}
    1 = \int \prod_{a,b} \diff Q_{a,b} \; \delta\left[Q_{a,b} - \frac{1}{\lambda_1 N_1}\sum_{\mu,\nu}^P \Bar{s}^\mu_{a} K_{\mu \nu}(C)\Bar{s}^\nu_{b} \right]
\end{equation}
and its dual $\bar Q_{a,b}$ via the Fourier representation of the deltas: 
 \begin{equation}
     \begin{split}
         Z =& \int  \diff \mathbf{Q} \bar{\mathbf{Q}} \;\det\left[\mathbb{1}_{\kappa}+\mathbf{Q}\right]^{-\frac{N_1}{2}} e^{i \sum_{a,b} Q_{a,b}\Bar{Q}_{a,b} } \int \prod_{a,\mu} \diff \Bar{s}_{a}^\mu \exp\left\{\frac{i}{\lambda_1 N_1}\sum_{a,b}\bar{Q}_{a,b}\sum_{\mu,\nu} \Bar{s}_{a}^\mu K_{\mu \nu} \bar{s}_{b}^\nu\right\}\\
         & \int \prod_{a,\mu} \diff s_{a}^\mu e^{-\frac{\beta}{2 \kappa}\sum_{a,\mu}\left(y_{a}^\mu-s_{a}^\mu\right)^2+i\sum_{a,\mu} s_{a,\mu}\bar{s}_{a,\mu}}.
     \end{split} 
 \end{equation}
In conclusion, the integrals in $s$ and $\bar{s}$ can be solved and we land with the following effective action $S$: 
\begin{equation}
\begin{aligned}
S(\mathbf{Q},\Bar{\mathbf{Q}})=& -\Tr[\mathbf{Q}\Bar{\mathbf{Q}}^{\top}]+\Tr\log(\mathbb{1}_{\kappa}+\mathbf{Q})+\frac{\alpha_1}{P}\Tr\log \frac{\beta}{\kappa}\left[ \frac{ \kappa}{\beta} \;\mathbb{1}_{\kappa}\otimes\mathbb{1}_P +\frac{ \bar{\mathbf{Q}}\otimes K}{\lambda_1}\right]+\frac{\alpha_1}{P} y^\top \left[ \frac{ \kappa}{\beta}\; \mathbb{1}_{\kappa}\otimes\mathbb{1}_P +\frac{ \bar{\mathbf{Q}}\otimes K}{\lambda_1}\right]^{-1} y.
\end{aligned}
\end{equation}

\section{1HL effective action and single output: special cases}
\label{supp:sec:special_cases}

In this section we report cases of activation functions for which we are able to evaluate analytically the probability distribution of the variable $q$, defined in the main text as
\begin{equation}
    q = \frac{1}{\sqrt{\lambda N_1}} \sum_{\mu=1}^P \bar{s}^\mu \sigma(h^\mu)\,,
    \label{supp:eq:q_def}
\end{equation}
at fixed instance of the vector $\bar{s}$, clarifying the conditions to impose on the data for $q$ to be Gaussian. Its characteristic function is defined as
\begin{equation}
    \psi(t) = \expect_q\left\{ \exp (iqt) \right\} =\expect_h\left\{ \exp\left[\frac{it}{\sqrt{\lambda N_1}} \sum_\mu \bar{s}^\mu \sigma(h^\mu)\right] \right\}\,.
    \label{supp:eq:characteristic_q}
\end{equation}
If $q$ is Gaussian, then $\psi = \phi$, where
\begin{equation}
    \phi(t) = \exp\left(-\frac{t^2 Q}{2} \right) = \exp\left( -\frac{t^2}{2 \lambda N_1} \sum_{\mu,\nu} \bar{s}^\mu K_{\mu\nu} \bar{s}^\nu \right)\,
    \label{supp:eq:characteristic_gauss}
\end{equation}
is the characteristic function of a Gaussian variable with variance given by
\begin{equation}
    Q = \frac{1}{\lambda N_1}  \sum_{\mu,\nu} \bar{s}^\mu K_{\mu\nu} \bar{s}^\nu\,, \qquad K_{\mu\nu} = \expect_h[\sigma(h^\mu) \sigma(h^\nu)]\,.
\end{equation}

\subsection{Linear activation function: \texorpdfstring{$q$}{q} is Gaussian at finite \texorpdfstring{$P$, $N_0$, $N_1$}{P, N0, N1}}

The case of $\sigma = \id$ has been already worked out in the literature, see~\cite{SompolinskyLinear, hanin2023bayesian}. We report it here for reference, and to stress that our theory reduces to known cases as it should. Indeed, when the activation function is linear the average over $h$ in Eq.~\eqref{supp:eq:characteristic_q} can be computed exactly at finite $P$, $N_1$, and gives
\begin{equation}
    \psi_{\text{lin}}(t) = \exp\left( -\frac{t^2}{2 \lambda N_1} \sum_{\mu,\nu} \bar{s}^\mu C_{\mu\nu} \bar{s}^\nu \right)\,.
\end{equation}
Note that $C$ is the value of the kernel for $\sigma = \id$. This is strictly true as long as $C$ has no zero eigenvalue, so at least for $N_0>P$; however, a small regularization proportional to the identity matrix can be added to $C$ to avoid this problem. 

This result is simply due to the fact that the sum of jointly Gaussian variables is Gaussian, which is true for generic Gram matrices $C$ and any value of $P$, $N_1$, even far from the asymptotic limit $P\sim N_1$ large. In order to evaluate the remaining integrals over the order parameters at the saddle-point of the effective action, this limit is still required, as performed indeed in~\cite{SompolinskyLinear}, to which our theory reduces; otherwise, for $P$, $N_1$ finite one can express the partition function exactly in terms of Meijer G-functions, see~\cite{hanin2023bayesian}.

\subsection{Quadratic activation function}
Let us take now $C_{\mu\mu} = 1$ (normalized data) and quadratic (zero-mean) activation function:
\begin{equation}
    \sigma(x) = x + a(x^2-1)\,.
\end{equation}
The kernel is given by
\begin{equation}
    K_{\mu\nu} = \expect_h[\sigma(h^\mu)\sigma(h^\nu)] =  C_{\mu\nu}  +2 a^2  (C_{\mu\nu})^2\,.
\end{equation}
Also in this case the characteristic function in~\eqref{supp:eq:characteristic_q} can be evaluated exactly:
\begin{equation}
\begin{aligned}
    \psi_{\text{quad}}(t) = \frac{\exp\left\{-\frac{t^2}{2\lambda N_1}  \bar{s}^\top C \left[\mathbb{1}_P - \frac{2i a t }{\sqrt{\lambda N_1}}\diag(\bar{s}) C \right]^{-1} \bar{s}\right\}}{\det[(\mathbb{1}_P - \frac{2 i a t }{\sqrt{\lambda N_1}}\diag(\bar{s}) C)]^{1/2}} \exp\left(- \frac{i a t}{\sqrt{\lambda N_1}} \sum_\mu 
    \bar{s}^\mu \right)\,.
    \label{supp:eq:quad_psi_exact}
\end{aligned}
\end{equation}
We can express the non-trivial matrices appearing in this expression as Neumann series:
\begin{align}
     \left[\mathbb{1}_P - \frac{2i a t }{\sqrt{\lambda N_1}}\diag(\bar{s}) C \right]^{-1} &=  \sum_{n=0}^{+\infty} \left(\frac{2i a t }{\sqrt{ \lambda N_1}}\right)^n [\diag(\bar{s}) C]^n \,,\label{supp:eq:quad_firstSeries}\\
     -\frac{1}{2} \Tr \log\left[\mathbb{1}_P - \frac{2i a t }{\sqrt{\lambda N_1}}\diag(\bar{s}) C \right] &= -\sum_{n=1}^{+\infty} \frac{1}{n} \left(\frac{2i a  }{\sqrt{\lambda N_1}}\right)^n \Tr\{[\diag(\bar{s}) C]^n\}\,.
     \label{supp:eq:quad_secondSeries}
\end{align}
To prove Gaussianity, we need to require the following asymptotic behaviors:
\begin{align}
    \frac{1}{N_1^{1+n/2}} \bar{s}^\top C  [\diag(\bar{s}) C]^n \bar{s} &= O(P/N_1^{1+n/2})\,, \label{supp:eq:quad_firstCondition} \\
    \frac{1}{N_1^{n/2}} \Tr\{[\diag(\bar{s}) C]^n\}  & = O(P/N_1^{n/2})\,, \label{supp:eq:quad_secondCondition}
\end{align}
so that in the regime where $\alpha_1 = P/N_1$ is finite only the $n=0$ term counts for~\eqref{supp:eq:quad_firstSeries} and the $n=1,2$ terms for~\eqref{supp:eq:quad_secondSeries}. Using
\begin{equation}
   -\frac{1}{2} \Tr \log \left[\mathbb{1}_P - \frac{2i a t }{\sqrt{\lambda N_1}}\diag(\bar{s}) C \right] \approx \frac{i a t }{\sqrt{\lambda N_1}}\sum_\mu \bar{s}^\mu - \frac{a^2 t^2 }{\lambda N_1} \sum_{\mu,\nu} \bar{s}_\mu  (C_{\mu\nu})^2 \bar{s}_\nu\,,
\end{equation}
we get
\begin{equation}
     \psi_{\text{quad}}(t) \sim  \exp\left[-\frac{t^2}{2 \lambda N_1} \sum_{\mu,\nu} \bar{s}^\mu K_{\mu\nu} \bar{s}^\nu \right]\,.
\end{equation}

The conditions~\eqref{supp:eq:quad_firstCondition},~\eqref{supp:eq:quad_secondCondition} should be interpreted as hypothesis on the Gram matrix of the data $C$ and on the realization of the vector $\bar{s}$ in order for the property of Gaussianity to hold. Let us see the simplest case of i.i.d. standard normal input data and $\bar{s}^\top = (1,\cdots,1)$. Then, $C$ is a Wishart matrix with a finite spectrum in the regime $P\sim N_0$~\cite{marchenko1967}, and
\begin{equation}
   \frac{1}{P}\Tr(C^n) = O(1)\,,\qquad\frac{1}{P}\sum_{\mu,\nu} (C^{n})_{\mu\nu} = O(1) \,.
   \label{supp:eq:quad_hypothesis}
\end{equation}
The first behaviour follows from the fact that the eigenvalues are $O(1)$, while the second can be proven using
\begin{equation}
    C = O(1) \mathbb{1}_P + O(1/\sqrt{N_0}) H\,,
\end{equation}
where $H$ is a symmetric random matrix with elements $\pm 1$, or, more formally, exploiting the fact that the eigenvectors of a Wishart matrix are random and uniformly distributed on the sphere~\cite{forresterBook}, so that
\begin{equation}
    \frac{1}{P}\sum_{\mu,\nu} (C^{n})_{\mu\nu} = \frac{1}{P}\sum_{\rho} \lambda^{n}_\rho \sum_\mu U_{\mu\rho} \sum_\nu U^{-1}_{\rho \nu} = O(1)\,,
\end{equation}
where $\lambda_\rho$ is the $\rho$-th eigenvalue of $C$ and $U$ the matrix whose $\rho$-th column is the corresponding eigenvector. Given that, properties~\eqref{supp:eq:quad_firstCondition},~\eqref{supp:eq:quad_secondCondition} follow and $q$ is Gaussian.

In principle, Gaussianity can be also proven via diagrammatic techniques. Take for example the quartic moment of the variable $q$ in~\eqref{supp:eq:q_def}. One can see, via Wick's theorem, that
    \begin{equation}
    \begin{aligned}
        \expect_h[&\sigma(h^{\mu_1})\sigma(h^{\mu_2})\sigma(h^{\mu_3}) \sigma(h^{\mu_4}) ] - (K_{\mu_1 \mu_2} K_{\mu_3 \mu_4} +K_{\mu_1 \mu_3} K_{\mu_2 \mu_4} + K_{\mu_1 \mu_4} K_{\mu_2 \mu_3} ) = \\
        &
        16 a^4 
        (C_{\mu_1 \mu_2} C_{\mu_1 \mu_3} C_{\mu_2 \mu_4} C_{\mu_3 \mu_4}+
            C_{\mu_1 \mu_2} C_{\mu_1 \mu_4} C_{\mu_2 \mu_3} C_{\mu_3 \mu_4}+
            C_{\mu_1 \mu_3} C_{\mu_1 \mu_4} C_{\mu_2 \mu_3} C_{\mu_2 \mu_4}
        )\\
        &+4 a^2 
        (C_{\mu_1 \mu_2} C_{\mu_1 \mu_3} C_{\mu_2 \mu_4}+
        C_{\mu_1 \mu_2} C_{\mu_1 \mu_3} C_{\mu_3 \mu_4}+
        C_{\mu_1 \mu_2} C_{\mu_1 \mu_4} C_{\mu_2 \mu_3}+\\
        &\hphantom{{}+4 a^2 
        (}
        C_{\mu_1 \mu_2} C_{\mu_1 \mu_4} C_{\mu_3 \mu_4}+
        C_{\mu_1 \mu_2} C_{\mu_2 \mu_3} C_{\mu_3 \mu_4}+
        C_{\mu_1 \mu_2} C_{\mu_2 \mu_4} C_{\mu_3 \mu_4}+\\
        &\hphantom{{}+4 a^2 
        (}
        C_{\mu_1 \mu_3} C_{\mu_1 \mu_4} C_{\mu_2 \mu_3}+
        C_{\mu_1 \mu_3} C_{\mu_1 \mu_4} C_{\mu_2 \mu_4}+
        C_{\mu_1 \mu_3} C_{\mu_2 \mu_3} C_{\mu_2 \mu_4}+\\
        &\hphantom{{}+4 a^2 
        (}
        C_{\mu_1 \mu_3} C_{\mu_2 \mu_4} C_{\mu_3 \mu_4}+
        C_{\mu_1 \mu_4} C_{\mu_2 \mu_3} C_{\mu_2 \mu_4}+
        C_{\mu_1 \mu_4} C_{\mu_2 \mu_3} C_{\mu_3 \mu_4}
        )\,,
    \end{aligned}
    \label{supp:eq:quad_diagrams0}
    \end{equation}
    while the quartic term from~\eqref{supp:eq:characteristic_gauss} involves only the diagrams
    \begin{equation}
    \begin{aligned}
        K_{\mu_1 \mu_2} K_{\mu_3 \mu_4}& +K_{\mu_1 \mu_3} K_{\mu_2 \mu_4} + K_{\mu_1 \mu_4} K_{\mu_2 \mu_3} =\\
        &4 a^4 (C_{\mu_1 \mu_2}^2 C_{\mu_3 \mu_4}^2+C_{\mu_1 \mu_3}^2 C_{\mu_2 \mu_4}^2+C_{\mu_1 \mu_4}^2 C_{\mu_2 \mu_3}^2)\\
        &+2 a^2 (C_{\mu_1 \mu_2}^2 C_{\mu_3 \mu_4}
        +C_{\mu_1 \mu_2} C_{\mu_3 \mu_4}^2
        +C_{\mu_1 \mu_3}^2 C_{\mu_2 \mu_4}
        +C_{\mu_1 \mu_3} C_{\mu_2 \mu_4}^2
        +C_{\mu_1 \mu_4}^2 C_{\mu_2 \mu_3}
        +C_{\mu_1 \mu_4} C_{\mu_2 \mu_3}^2)\\
        &+C_{\mu_1 \mu_2} C_{\mu_3 \mu_4}+C_{\mu_1 \mu_3} C_{\mu_2 \mu_4}+C_{\mu_1 \mu_4} C_{\mu_2 \mu_3}\,.
    \end{aligned}
     \label{supp:eq:quad_diagrams1}
    \end{equation}
    This is not surprising: the variables $\sigma(h^\mu)$ are not Gaussian due to the non-linearity. However, when summed over all the indices, the diagrams in Eq.~\eqref{supp:eq:quad_diagrams0}
    are of the form $\Tr C^4 $ or $\sum_{\mu,\nu} (C^3)_{\mu \nu} $, both $O(P)$ under the hypothesis stated above, while the 
    diagrams in~\eqref{supp:eq:quad_diagrams1} are of the form $(\sum_{\mu,\nu} (C^2)_{\mu\nu})^2$, $(\sum_{\mu,\nu} C_{\mu\nu})^2$ or $(\sum_{\mu,\nu} (C^2)_{\mu\nu})(\sum_{\mu,\nu} C_{\mu\nu})$, which are all $O(P^2)$ and leading over the first ones.

    As long as $\bar{s}^\mu \sim O(1)$ for all $\mu$, we do not expect the previous derivation to change. On the other hand, we point out that there exist special configurations $\bar s$, such as $\bar s^\top =(1,0,\cdots,0) $, for which this reasoning breaks down. As such, we are assuming that the contribution of these special configurations to the effective action is negligible in the thermodynamic limit.

\section{Generalisation to deep neural networks with \texorpdfstring{$L$}{L} hidden layers: derivation of the saddle-point equations in special cases}
\label{supp:sec:LHL}

In the next sections we consider two cases where simplifications arise. These special cases correspond to kernels $K$ such that $K(\alpha C) = \alpha^s K(C)$, where $\alpha$ is any positive scalar and $s \geq 0$ is an integer. It turns out that $s=0$ is realized by the sign activation function, whereas $s=1$ holds for piece-wise linear activations such as ReLU or Leaky-ReLU.

\subsection{Saddle-point equations for scale independent kernels of the form \texorpdfstring{$K(\alpha C) = K(C)$ ($\alpha>0$)}{K(alpha C) = K(C)}}

In the case of sign activation function, it is straightforward to show that the behavior under scalar multiplication of the kernel $K^{(R)}_L(C)$ follows from the property $\textrm{sign} (\alpha x) = \textrm{sign} (x)$. It turns out that in this special case the effective action for deep learning considerably simplifies, since the non-linear dependence of $K^{(R)}_L$ on the variables $\{\bar Q_\ell\}_{\ell\neq L}$ disappears. This allows to solve the saddle-point equations exactly. In particular:
\begin{equation}
Q_\ell^* = 0\,, \quad \bar Q_\ell^* = 1\, \quad \forall \, \ell=1,\dots,L-1\,,
\end{equation}
whereas the functional form of the solution for $\bar Q_L$ is the same as in the one-hidden layer case (in the zero temperature limit):
\begin{equation}
\bar Q_L^* = \frac{\sqrt{(\alpha_L-1)^2 + 4 \alpha_L \frac{1}{P} y^T K_L^{-1}y}-(\alpha_L -1) }{2}\,.
\end{equation}
In practice, such a solution shows that deep architectures with $\textrm{sign}$ activation (that are problematic to employ in practice since it is challenging to backpropagate derivatives) essentially behave as one hidden layer neural networks in the proportional limit and the only marker of the depth $L$ is retained in the infinite-width kernel $K_L$.

\subsection{Saddle-point equations for piecewise linear kernels of the form \texorpdfstring{$K(\alpha C) = \alpha K(C)$}{K(alpha C) = alpha K(C)}}
\label{saddle:ReLu}

The linear behavior of the kernel under scalar multiplication follows for ReLU and leaky ReLU activation function from the property $\textrm{ReLU} (\alpha x) = \alpha\, \textrm{ReLU} (x)$. It turns out that in this case the effective action reads:
\begin{align}
S_{\textrm{DNN}}(\{Q_\ell, \bar Q_\ell\}) &= \sum_{\ell=1}^{L} \frac{\alpha_L}{\alpha_\ell}\left[-Q_\ell \bar Q_{\ell} + \log(1+Q_\ell)\right] +\frac{\alpha_L}{P}\text{Tr}\log \beta \left[  \frac{\mathbb{1}}{\beta} +\left(\prod_{\ell=1}^L\bar{Q}_\ell\right) K_L(C)\right] \notag\\
&\quad+\frac{\alpha_L}{P} y^T \left[ \frac{\mathbb{1}}{\beta} +\left(\prod_{\ell=1}^L\bar{Q}_\ell\right) K_L(C)\right]^{-1} y\,.
\end{align}

Exactly as for the one hidden layer case, the saddle-point equations simplify in the zero temperature limit and under the assumption that the $L$-hidden layers kernel $K_L$ has only positive eigenvalues:
\begin{equation}
Q_\ell \bar Q_\ell
- \alpha_\ell \\
+\frac{\alpha_\ell}{\left(\prod_{\ell_1} \bar Q_{\ell_1}\right)}\frac{1}{P} y^T K_L^{-1} y = 0
\label{saddleReLU}
\end{equation}
for all $\ell=1,\dots,L$.

Notice that if $\alpha_\ell = \alpha$ for all $\ell=1, \dots, L$, it is easy to show that the only solution must satisfy $Q_\ell^* = Q^*$ for all $\ell$ and we recover the heuristic mean field theory proposed in Ref. \cite{SompolinskyLinear}. The reason for this equivalence is obvious: the authors of \cite{SompolinskyLinear} found the heuristic mean field theory for ReLU activation by replacing the linear kernel with the corresponding NNGP kernel, noticing that the ReLU kernel transforms as the linear one under multiplication by a scalar. Our derivation shows that this replacement is not correct for general activation functions (see for instance the case of sign activation previously discussed), but it is possible in this particular case.

For completeness, we also show how to re-derive the self-consistent equations found by Li and Sompolinsky \cite{SompolinskyLinear} in the linear case. The effective action for the linear case reads:
\begin{align}
S_{\textrm{DNN}}(\{Q_\ell, \bar Q_\ell\}) &= \sum_{\ell=1}^{L} \frac{\alpha_L}{\alpha_\ell}\left[-Q_\ell \bar Q_{\ell} + \log(1+Q_\ell)\right] +\frac{\alpha_L}{P}\text{Tr}\log \beta \left[  \frac{\mathbb{1}}{\beta} +\left(\prod_{\ell=1}^L\bar{Q}_\ell\right) C_L\right] \notag\\
&\quad+\frac{\alpha_L}{P} y^T \left[ \frac{\mathbb{1}}{\beta} +\left(\prod_{\ell=1}^L\bar{Q}_\ell\right) C_L\right]^{-1} y\,,
\end{align}
where $C_L = C/(\prod_{\ell=1}^L \lambda_\ell)$ and $C_{\mu\nu} = \mathbf x^\mu\cdot \mathbf{x}^{\nu}/(\lambda_0 N_0)$. Let us consider the case of isotropic aspect ratios $\alpha_\ell = \alpha$, $\forall \ell=1,\dots, L$ and same Gaussian priors at each layer $\lambda_\ell = \lambda$, $\forall \ell=0,\dots,L$. The saddle-point equations for $\bar Q_\ell$ read: 
\begin{equation}
    1-\bar Q_\ell = \alpha\left(1-\frac{\lambda^L}{\left(\prod_{\ell_1} \bar Q_{\ell_1}\right)}\frac{1}{P} y^T C^{-1} y\right)\,.
\end{equation}
It turns out that we recover the equation for the renormalization parameter $u_0$ in \cite{SompolinskyLinear} by noticing that the only solution of this system of equations is of the form $\bar Q_\ell = \bar Q^*$ and by making the identification $u_0 = \bar Q^*/\lambda$.

\section{Generalising the effective action to finite-mean activation functions}
\label{supp:sec:finitemeanactivation}

In this section we show how the theory can be generalized in the case of finite-mean activation functions. In fact, up to this point, our derivation assumed that the integral of the activation function over a centered Gaussian is zero, i.e. the activation function is zero-mean. The goal of this section is to show that removing such hypothesis modifies the effective action in the asymptotic limit. Since ReLU activation belongs to this more general case, the findings of this section imply that Li-Sompolinsky heuristic theory \cite{SompolinskyLinear} should be modified as well. As for the rest of the manuscript, we start by considering one hidden layer architectures and we later extend the result to $L$ hidden layers.

The crucial difference wrt to the case studied in section \ref{sec:1HL} is that if the activation function is not zero-mean, also the random variable 
\begin{equation}
q = \frac{1}{\sqrt {N_1 \lambda_1}} \sum_{\mu=1}^P \bar s^\mu \sigma(h^\mu) 
\end{equation}
has now a finite mean. In particular:
\begin{align}
\braket{q}_{P(q)} = \frac{1}{\sqrt {N_1 \lambda_1}} \sum_{\mu=1}^P \bar s^\mu m^{\mu}\,,  \qquad
m^{\mu} = \int \frac{dt}{\sqrt{2\pi C_{\mu\mu}}} e^{- t^2/ (2C_{\mu\mu})} \sigma(t) \,.
\end{align}
A straightforward calculation shows that the result for finite-mean activation is found by performing the replacement:
\begin{align}
\frac{\bar Q}{\lambda_1} K \to  K^{(R)}(Q,\bar Q) =  \frac{\bar Q}{\lambda_1} K - \frac{\left(\bar Q - \frac{1}{1+Q} \right)}{\lambda_1} K^{(1)}\,, \qquad K^{(1)}_{\mu\nu} = m^{\mu} m^{\nu}\,
\end{align}
in the effective action in Eq. \eqref{effS} of the main text. As such, the one-hidden layer action for finite-mean activation functions reads:
\begin{equation}
\begin{aligned}
S_{\textrm{1HL}}= {}& -Q\bar{Q}+\log(1+Q)+ +\frac{\alpha_1}{P}\Tr\log \beta\left[ \frac{ \mathbb{1}}{\beta} +K^{(R)} (Q, \bar Q)\right]+\frac{\alpha_1}{P} y^\top \left[ \frac{\mathbb{1}}{\beta} +K^{(R)} (Q, \bar Q)\right]^{-1} y
\end{aligned}
\label{supp:effSfinitemean}
\end{equation}
It is worth noticing that while in the zero mean case there was a simple relations between $Q$ and $\bar Q$ at any temperature, we now lose that property and the saddle-point equations are not exactly solvable anymore, not even in the zero temperature limit. On the contrary, one can check that in the infinite-width limit we recover the previous result $\bar Q = 1$, $Q = 0$ and the rank one matrix $K^{(1)}$ does not contribute to the generalization error, since it does always appear in combination with the scalar $\bar Q - 1/(1+Q)$ that vanishes in the infinite-width limit.

Let us move to the derivation of an effective action for $L$ hidden layers. As for the derivation with zero-mean activation function, the key step is to understand how the joint probability of the pre-activations at layer $\ell$ is linked to the one at layer $\ell -1$. While in the zero-mean activation case, the key observation was that $P_2$ is related to $P_1$ by the replacement $C \to \bar Q_1 K (C)/\lambda_1$ (see Eq. \ref{preact12}), here we find that the correct replacement is $C \to \bar  Q_1 K (C)/\lambda_1-(\bar Q_1 - 1/(1+Q_1))K^{(1)}/\lambda_1$. Differently from the zero-mean activation case, where the kernel at layer $L$ was only depending on the variables $ \{\bar Q_\ell\}$, here we find that the recurrence is given in terms of the $\{Q_\ell\}$ as well. In conclusion, this produces a more unpleasant action where all the $\{Q_\ell,\bar Q_\ell\}$ are coupled via the nested non-linear expression of the kernel. The explicit recurrence relation for finite-mean activation functions is given by:
\begin{equation}
\begin{split}
K^{(R)}_{\ell} = \frac{\bar Q_{\ell}}{\lambda_{\ell}} K \circ \left[K^{(R)}_{\ell-1}\right] - \frac{\left(\bar Q_\ell -\frac{1}{1+Q_\ell}\right)}{\lambda_{\ell}}K^{(1)} \circ \left[K^{(R)}_{\ell-1}\right]\,, \qquad K^{(R)}_0 = C\,
\end{split}
\end{equation}
and the effective saddle-point action reads:
\begin{align}
S_{\textrm{DNN}}= \sum_{\ell=1}^{L} \frac{\alpha_L}{\alpha_\ell}\left[ -Q_\ell \bar Q_{\ell} + \log(1+Q_\ell)\right] +\frac{\alpha_L}{P}\text{Tr}\log \beta \left( \frac{\mathbb{1}}{\beta} +K^{(R)}_L( \{\bar Q_\ell, Q_\ell\})\right)+\frac{\alpha_L}{P} y^T \left( \frac{\mathbb{1}}{\beta} + K^{(R)}_L( \{\bar Q_\ell,Q_\ell\})\right)^{-1} y\,.
\label{supp:Eq:Sfinitemean}
\end{align}

In view of the above considerations, it should be now clear that the heuristic Li-Sompolinsky theory (re-derived in the previous section) amounts to disregard all the additional terms $K^{(1)}$ that arise from the approach presented in this section. 

\section{Numerical issues in sampling from the Bayesian posterior}
\label{supp:sec:numerical_issues}
Obtaining a perfect agreement between theory and simulations when sampling from a Bayesian posterior (especially in the zero temperature limit) is prevented by a number of technical numerical issues presented in the following.

\begin{itemize}
    \item[1.] Finite-size effects certainly play a role in explaining the small mismatch between theory and experiment. To address this point, we are currently performing high-precision numerical simulations with fixed $\alpha = P /N_1$ and increasing values of $N_1$ and $P$.
    \item[2.] The $T \to 0$ limit, which corresponds to perfect interpolation of the dataset and is the only case in which the saddle point equations can be solved analytically, was the most logic to address for starting, but turns out to be very hard to simulate. This is clear from some preliminary work we are doing, where we numerically solve the saddle point equations at generic $T$ for the saddle point variables $Q $,  $\bar Q = f(Q)$. We find that the function $Q(T)$ changes rapidly for small temperatures. 
    \item[3.] At $T= 0.001$, the autocorrelation time of the simulation is already very large, taking as little as $5 \cdot 10^6$ epochs to thermalize. As the temperature is decreased, the autocorrelation time increases, and we need hundreds of thousands of epochs to gain satisfactory statistics. 
    \item[4.] The effect of a finite learning rate $\eta$ has to be taken into account as well. From our preliminary results, we empirically observe that finite-$\eta$ effects are larger at higher temperature. The standard way to take into account finite-$\eta$ effects is to perform the extrapolation to $\eta \to 0$ simulating different learning rates. 
	\item[5.] Computing the theory in the case of $L > 1$ networks requires to numerically minimize a complex nested saddle-point functional of the variables $\bar Q_{\ell}$. We are currently working on a numerical routine to efficiently perform this task.
\end{itemize}

\end{document}